\documentclass[12pt, titlepage]{article}%
\usepackage{graphicx}
\usepackage{amsmath}
\usepackage{amsfonts}
\usepackage{amssymb}
\usepackage{fancyvrb}
\usepackage[doublespacing]{setspace}
\providecommand{\U}[1]{\protect\rule{.1in}{.1in}}
\def\baselinestretch{1.5}

\usepackage{color}

\begin{document}

\title{Swimming with Wealthy Sharks: \\ Longevity, Volatility and the Value of Risk Pooling}
\author{Moshe A. Milevsky\thanks{Milevsky is professor of finance at the Schulich School of Business and a member of the graduate faculty of Mathematics and Statistics at York University. He can be reached at: milevsky@yorku.ca, or at Tel: 416-736-2100 x 66014. His address is: 4700 Keele Street, Toronto, Ontario, Canada, M3J 1P3. The author would like to acknowledge: Michael Stepner for access and help with U.S. mortality data, Kevin Milligan for access and help with Canadian mortality data, research assistance from Victor Le, Dahlia Milevsky and Yossef Bisk, editorial assistance from Alexa Brand and Alexandra Macqueen, encouraging comments from the inseparable Natalia Gavrilova and Leonid Gavrilov, suggestions from discussant Lars Stentoft and participants at the CEAR-RSI Household Finance Workshop in Montreal (November 2018), feedback from two anonymous JPEF reviewers and many conversations about (modeling) life and death with Tom Salisbury and Huaxiong Huang.}}
\date{22 November 2018}
\maketitle

\begin{abstract}

\begin{singlespace}
\begin{center}
{\bf Swimming with Wealthy Sharks: \\ Longevity, Volatility and the Value of Risk Pooling }
\end{center}
\end{singlespace}

\begin{singlespace}
Who {\em values} life annuities more? Is it the healthy retiree who expects to live long and might become a centenarian, or is the unhealthy retiree with a short life expectancy more likely to appreciate the pooling of longevity risk? What if the unhealthy retiree is pooled with someone who is much healthier and thus forced to pay an implicit loading? To answer these and related questions this paper examines the empirical conditions under which retirees benefit (or may not) from longevity risk pooling by linking the {\em economics} of annuity equivalent wealth (AEW) to {\em actuarially} models of aging. I focus attention on the {\em Compensation Law of Mortality} which implies that individuals with higher relative mortality (e.g. lower income) age more slowly and experience greater longevity uncertainty. Ergo, they place higher utility value on the annuity. The impetus for this research today is the increasing evidence on the growing disparity in longevity expectations between rich and poor.
\end{singlespace}

\vspace{0.5in}

\begin{singlespace}
\begin{flushleft}
\end{flushleft}
\end{singlespace}

\end{abstract}

\clearpage
\begin{flushright}
{\em There's nothing serious in mortality, \\
all is but toys, renown and grace is dead.} \\
{\bf Macbeth, Act II, Scene III}
\end{flushright}
\section{Background and Motivation}

The noted Princeton economist and Nobel laureate Angus Deaton recently wrote that: ``The finding that income predicts mortality has a long history,'' having been noted as far back as the 19th century by Friedrich Engels in Manchester, England. According to Deaton (2016), commenting on similar findings by Chetty et al. (2016), ``There is little surprising in yet another study that shows that those with higher income can expect to live as much as 15 years more than those with lower income." It simply isn't news. Indeed, the focus among those (economists) who study \textit{mortality and its inequality}, using a phrase coined by Peltzman (2009), has shifted to the causes and consequences as opposed to proving its existence. The question en vogue is: {\em Why is the mortality gradient steepening?} and {\em Why is it worse in some countries versus others?}

What has received less attention from economists -- and in fact may be surprising to many -- is that not only is the longevity of those at the lowest income percentiles in the U.S. lower, the uncertainty or variability of their remaining lifetime is higher as well. It is the exact opposite of the well-known relationship in portfolio theory. If one thinks of the remaining lifetime random variable: $T_x$ in terms of {\em return} (i.e. expected length) and {\em risk} (i.e. dispersion), then generally speaking the mean value is lower but the standard deviation is higher for individuals with lower income (i.e. poorer), but at the same chronological age. And, when these two numbers are expressed as a ratio (a.k.a. the coefficient of variation of longevity, or CoVoL) the difference is even more pronounced. 

The positive (cor)relation between mortality rates and the volatility of longevity (as well as standard deviation) for individuals follows from the {\em Compensation Law of Mortality} (CLaM). This theory was introduced in 1978 (in Russian) and expanded on by Gavrilov and Gavrilova (1991) in their textbook. I'll elaborate on this later on, but to be clear, it's a theory and remains somewhat controversial. At this point all I'll say is that if you are unfortunate enough to have high mortality you also face higher longevity uncertainty.

Now, more than a statistical curiosity or something to idly puzzle over, nowhere is the natural link between life expectancy and its volatility more pertinent than in the area of pensions, retirement planning and (subjective) annuity valuation.

\subsection{Pension Subsidies}

For the sake of a wider readership (but at the acknowledged risk of alienating an academic audience) I'll dispense with tradition and motivate this paper with a very simple example. Assume that Mrs. Heather is about to retire at the age of 65 and is now entitled to a pension annuity, a.k.a. social security, or guaranteed lifetime income of exactly \$25,000 per year, paid monthly. For the record, this is the legislated maximum she can receive after having worked the requisite number of years. In other words, she has also contributed the maximum, perhaps explicitly by having a fraction of her paycheck withheld or implicitly via the income tax system. The pension annuity payments are adjusted for the cost of living or price inflation as measured by some national index, but the income will cease upon her death. The pension annuity contains no cash value or liquidity provisions, nor can she bequeath the income to her children, grandchildren or loved ones. I'm not describing any one specific country or government plan, but rather a generic, no-frills defined benefit (DB) pension scheme managed by any large sponsor.

Coincidentally, her next-door neighbor Mr. Simon was born in the same year, is also about to retire at 65 and is entitled to the same annuity of \$25,000 per year. He, too, has contributed the maximum to the scheme. The exact details of how Simon and Heather paid for their pension annuity are unimportant at this juncture. The key is that over his working life -- and in particular by the time he retires -- both Simon and Heather have contributed fully to the pension system.

There is one important difference between the two, though. Regrettably, Simon has a medical life expectancy of 10 years and is rather sickly, whereas Heather is in perfect health with a corresponding life expectancy of 30 years -- and they both know it. Heather is expected live to age 95 (from her current age of 65) and outlive Simon who will only make it to 75.  Stated differently, although their chronological ages (CA) are both 65, Heather's biological age (BA) is much lower than Simon's BA. In actuarial terms, his mortality hazard rate is substantially higher than Heather's. 

And yet, despite Simon's poor health and the financial fact he contributed the exact same amount to the retirement program or pension scheme, he isn't entitled to any more income from his pension scheme's annuity than Heather. In the language of insurance, retirement programs aren't underwritten or adjusted for individual health status. Instead, {\em all} government social security programs around the world are unisex or gender neutral. If you contribute (the same amount) into the system, you receive the same benefit regardless of your sex, health status or any other bio-marker for longevity.

Obviously, Simon's shorter life expectancy of ten years implies that he will be receiving (much) less money back compared to Heather. Moreover, if the retirement program or scheme is designed to neither make or lose money in the long run, a.k.a. it is actuarially balanced, then the (sick) Simons of the world are subsidizing the (healthy) Heathers. Economists know this very well and it is the nature of all government pension programs. In fact, in most of Europe today, insurance companies are prohibited from using gender to price {\em any} type of retail policy, whether it be life, health, home or even car insurance. (In other words, Heather has to pay a bit more for car insurance, in Europe, relative to Simon.)

My objective in introducing this very simple framework is to quantify the magnitude of the financial subsidy from Simon to Heather, one that will form the basis and intuition for what follows later. Of course, despite the large dollar value of the transfer, the entire point of this paper is to argue why (and quantify how) Simon might still benefit from being a member of a pension scheme due to the fact that his longevity risk is greater. 

To properly analyze the subsidy from a financial perspective, the natural next step is as follows: How much would Simon have to pay in the open (a.k.a. retail) market to acquire a pension annuity of \$25,000 per year, and how much would Heather have to pay? That price or cost should give a rough sense of the magnitude of the transfer provided by Simon to Heather. 

In practice, the market price will depend on many loading factors and, more importantly, the magnitude of the uncertainty around Simon and Heather's life expectancy. But to keep things very simple (at this early intuitive stage) I'll assume that Simon will live for exactly 10 years and Heather will live for exactly 30 years. In other words, remaining lifetimes are deterministic and Simon would be charged \$212,750 at the chronological age of 65 and Heather would be charged \$487,250 for the same exact (term-certain) annuity. 

These numbers are based on a 3\% effective (real) annual interest rate but do not require much else in terms of assumptions or parameters. Stated differently, the present value of \$25,000 per year for 10 years is exactly \$212,750, for 30 years of cash flows, the present value is \$487,250. Moreover, the market cost of their combined pension annuity entitlement is \$212,750 + \$487,250 = \$700,000, a very important number from a funding and pension solvency perspective. So, if -- and this is a big if -- the pension scheme is actuarially balanced,  it should have exactly \$700,000 set aside in reserves to pay pension annuities when Heather and Simon both retire\footnote{Note: Few schemes have anywhere near \$700,000 set aside to pay all the guaranteed pension annuities under reasonable discount rates. For the most prominent voice in this area, see Joshua Rauh.}.

To be clear, real-world insurance companies will not charge \$212,750 and \$487,250 to Simon and Heather for these annuities. First and foremost, these companies have to make profits, so they would mark up or ``load" the price, just like the retail vs. wholesale cost of coffee. More importantly, insurance companies have to budget and provision for uncertainty, including the risk of how long their annuitants will live. I'll get to more refined mortality models in section \#3. 

The relevant upshot is as follows: Simon is transferring \$137,250 to Heather -- a subsidy amounting to 64.5\% of the hypothetical value of Simon's pension pot. Where did this number come from? Again, the entire system should have \$700,000 set aside for both of them, of which \$487,250 is needed for Heather and only \$212,750 is required for Simon. And yet, by construction they both contributed the same amount of money to the pension scheme, which presumably is a total of \$350,000 x 2 = \$700,000 over the course of their lives. To repeat, Simon contributed \$350,000 and is getting something worth \$137,250 {\em less} in the open market. Heather contributed \$350,000 and is getting something worth \$137,250 {\em more} in the open market. See Table \#1.

\begin{center}
\fbox{TABLE \#1 GOES HERE}
\end{center}

The numbers assume a simplistic pension system with no survivor benefits and an (extreme) 20-year gap in life horizon between the two (and only two) participants. More importantly I assumed Simon and Heather die precisely at their life expectancy, which presumes the absence of any longevity uncertainty -- which is the driver of insurance utility. 

In fact, Simon might live beyond age 75 (or he may die even earlier) and Heather might not make it to age 95 (or she may live even longer). Under these {\em ex-post} outcomes the cross subsidy will be smaller (or perhaps even larger) and hints at the insurance aspects of these schemes, something I'll return to in a moment.

But, the {\em ex-ante} reality is that there is a large gap between the expected present value of the benefits they receive even though they have paid similar amounts into the retirement program. In fact whenever you mix (a.k.a. pool) heterogenous people with different longevity prospects into one scheme in which everyone gets a pension annuity for the rest of their life, there will be winners and losers {\em ex ante} as well as {\em ex post}. This outcome is well-known in the pension and insurance economics literature\footnote{In fact this is a concern with {\em en vogue} Notional Defined Contribution (NDC) schemes, as noted recently by Holzman et al. (2017)}, but it is often surprising to non-specialists.

\subsection{Enter Longevity Risk}

What happens if we incorporate longevity risk or horizon uncertainty? Well, as I noted earlier there is a small probability that Simon lives for more than 10 years beyond age 75 and/or that Heather dies before age 95. Nobody really knows exactly how long they are going to live {\em ex ante}. In that case the {\em ex-post} transfer of wealth from Simon to Heather was less than 65\% of the value of his pension annuity. At the extreme edge there is a (very) small probability that Simon actually outlives Heather and the {\em ex-post} transfer goes in reverse; she subsidizes him. We won't know until all the Heathers and Simons are dead. 

Here is the main economic point. The pension annuity they are entitled to for the rest of their life provides them with more than just a periodic cash flow or income, it provides longevity insurance. Moreover, the value or benefit of any type of insurance can't be quantified in terms of what might happen {\em on average.} It must account for the tails of the distribution, which is best measured via (some sort of) discounted expected utility. 

Back to Heather and Simon. As mentioned, their annuity entitles them to more than a term-certain annuity for 30 and 10 years respectively -- they have acquired longevity insurance that protects them in the event they live longer. Simon would rather be {\em pooled} with people like him who share the same risk profile, as he would then expect a ``more equal" distribution. But even Simon is willing to be pooled with Heather if the alternative is no pooling at all. 

So, who values this insurance more? Heather or Simon? Or is the insurance benefit symmetric? Stated differently, could Simon be gaining more (in utility) from pooling with Heather, even if he is losing on an expected present value basis? The answer is yes, Simon could be winning (economic utility) even if he appears to be losing (dollars and cents). Why? Because his {\em volatility of longevity} is greater. Simon has a short life expectancy, but relatively speaking the range of how long he might actually live, both expressed in (i.) years: $SD[T_x]$, as well as (ii.) a fraction: $SD[T_x]/E[T_x]$, is actually greater than Heather's. 

If Simon lives 30 years instead of 10 years, that is equal to a 300\% (of mean lifetime) shock. It's unlikely, but in the realm of possibility. In contrast, Heather who is expected to live 30 years will never experience a 300\% shock. This would implies she lives 90 more years (from age 65) to the age of 155. It simply won't happen. The odds are zero. Ergo, Simon's individual volatility of longevity is higher than Heather's. That's displayed in Figure (\#1) and will be explained later.

\begin{center}
\fbox{FIGURE \#1 GOES HERE}
\end{center}

From an insurance economics point of view, it implies that Simon values the risk pooling benefits of the pension more than Heather does. The volatility of which I speak (and model) is more subtle than the likelihood of living 300\% longer than her current life expectancy. It will be defined precisely in section (\#3.) Most pension economists know that the transfer from Simon to Heather isn't as large as the expected dollars indicate, because (using my term) Simon's uncertainty or risk is larger than Heather's. He places greater value on the insurance component. More importantly, he is willing to swim in a {\em pool} with Heather, rather than taking his longevity chances.

Of course, Simon is a euphemism for unhealthy individuals who retire and aren't expected to live very long, whereas Heather represents retirees with long life expectancies. At this point I should make it clear that this isn't a matter of gender, race or nationality. There is a growing body of evidence that we can identify the Simons of the world {\em ex ante} based on the size of their wallets and magnitude of their income, not only via health or genetic testing. Nevertheless, forced risk pooling can (still) benefit everyone when measured in units of utility, and this is well known to pension and insurance economists, as the gap in life expectancy isn't too large and gap in uncertainty isn't too small. That's an empirical question.

The main contribution of this article is to leverage the data in Chetty et al. (2016), which documents an increasing gap in life expectancy based on income, to test if pooling is still valuable to Simon and Heather in the U.S. today, versus a few decades ago.\footnote{The same gap exists in Canada, per Milligan and Schirle (2018) but apparently isn't growing over time.}

We know today that at the chronological age of 50, taxpayers in the lowest income percentile (have much higher mortality rates and) are expected to live 10-15 years less than taxpayers in the highest income percentile. And yet, they are all forced to participate in the same (mandatory) social security program. Again, this paper uses the Chetty et al. (2016) mortality data (which I'll explain) to calibrate the extent of the transfer via the so-called {\em annuity equivalent wealth} in a Gompertz framework. To pre-empt my results, I show that even the lowest income percentiles in the U.S., still benefit from pooling because their volatility of longevity is large enough to overcome the implicit loading that comes from pooling. That is for the time being (today) and assuming they don't have any other sources of guaranteed lifetime income. More on this later.

\subsection{Outline of the Paper}

The remainder of this paper is organized (and presented in a more academic tradition) as follows. In the next section (\#2) I provide a proper literature review, linking the current paper to prior work in the field. Then, in section (\#3), I provide analytic context to the volatility or risk of longevity, by introducing the Gompertz law of mortality as well as the so-called {\em Compensation Law of Mortality}. Section \#4 provides an expression for the {\em annuity equivalent wealth} (AEW), which is another way of presenting the {\em willingness to pay} (WtP) for longevity insurance. I illustrate how it's a function of mortality characteristics and provide some examples. Section \#5, which is the main empirical contribution of this paper, displays and discusses the AEW as a function of income percentiles in the U.S. Section \#6 concludes the paper with the main takeaways. All tables and figures are located at the end of the document and technical derivations are relegated to an appendix.

\section{Scholarly Literature Review}

This paper sits within the so-called {\em annuity economics} literature, which attempts to explain the demand, or lack thereof, for insurance products that hedge personal longevity risk.  Life annuities are an important form of retirement income insurance, very similar to Defined Benefit (DB) pensions, as explained and advocated by Bodie (1990) for example. This literature began close to 50 years ago with the 1965 article by Menahem Yaari, in which he extended the standard lifecycle model to include actuarial notes. Yaari (1965) proved that for those consumers with no bequest motive, the optimal lifecycle strategy is to annuitize 100\% of assets. Clearly, few people have 100\% of their wealth annuitized (or ``pensionized") and even fewer actively purchase annuities, as pointed out by Franco Modigliani in his 1986 Nobel Prize address. 

The restrictive conditions in the original Yaari (1965) model were relaxed by Davidoff, Brown and Diamond (2005), and still the important role of annuities prevail. In fact, to quote the recent paper by Reichling and Smetters (2015), ``The case for 100\% annuitization of wealth is even more robust than commonly appreciated'' and it takes quite a bit of modeling effort to ``break'' (in their words) the Yaari (1965) result. Of course, including bequest and altruistic motives will reduce the 100\% annuitization result because annuity income dies with the annuitant and there is no legacy value. In a comprehensive review and modeling effort, Pashchenko (2013) pinpoints the extent to which bequest motives, pre-annuitized wealth and impediments to small annuity purchases can deter the full annuitization result. The (negative) impact of bequest motives is also echoed in work by Inkman, Lopes and Michaelidis (2010), who interestingly find a positive relationship between annuity market participation and financial education. 

Reichling and Smetters (2015) succeed in cracking the 50-year-old model by introducing stochastic mortality rates, in which the present value of the annuity is correlated with medical costs. According to them, although the annuity helps in protecting against the impact of longevity risk, its economic value is reduced in states of nature that are most costly to the retiree -- namely in the case of medical emergencies. In that state of nature annuities aren't as desirable; and as a result fewer people (than previously thought) should be acquiring any more annuities. 

Other attempts to ``break''  the Yaari (1965) model revolve around the underlying (additive) lifecycle model and moving-away from the implied risk neutrality over the length of life, towards a model with recursive preferences. See also Bommier (2006). To be clear, I don't aim for another crack in the Yaari (1965) model, or provide reasons for why consumers don't annuitize. 

The recent work in behavioral economics,  specifically the article by Brown et al. (2008), provides a rather convincing explanation for why consumers dislike annuities; having to do with framing, anchoring and loss aversion. The current paper stays well within the neoclassical paradigm, assuming that consumers are rational, risk-averse and maximizing an additive utility of consumption over a stochastic life horizon. 

This is the approach taken by Levhari and Mirman (1977), Davies (1981), Sheshinski (2007), or more recently Hosseini (2015), to name just a few. Moreover, I assume the consumer values annuities using the {\em annuity equivalent wealth} (AEW) metric introduced by Kotlikoff and Spivak (1981), also used by Brown (2001) and others who have calibrated these numbers around the world. The AEW is just another (reciprocal) way of presenting the {\em willingness-to-pay} metric, which is widely used in economics and recently reviewed in Barseghyan et al. (2018). Brown (2001) showed that an increase in the individual's AEW leads to an increase in the propensity to annuitize. It partially predicts who is likely to buy an annuity,\footnote{See also Brown, Mitchell, Poterba and Warshawsky (2001) and the link to money's worth ratio.} which is yet another reason to dig deeper in the structure of AEW.

That said, the main focus of attention in the current paper has to do with mortality {\em heterogeneity} and the subjective or personal value of annuities when everyone is forced to pay the same price, i.e., they all must swim in the same pool. 

Evidence of increasing mortality inequality continues to accumulate, and in particular the recent work by Chetty at al. (2016) indicates that the gap in expected longevity between the highest and lowest income percentiles in the U.S. can be as much as 15 years. These numbers are greater than (say) the 10 years reported by De Nardi, French and Jones (2009), or the 5-year gap noted in Poterba (2014, table 3) within the context of pensions and social security.  Peltzman (2009) notes that in the year 2002 U.S. life expectancy (by county) at the highest decile was 79.83 years, and at the lowest decile was 73.17 years, a gap of less than seven years. 

In contrast, Chetty et al. (2016) indicate that the measurable gap can extend to as much as 15 years. Echoing the same trend, within the context of social security, Goldman and Orszag (2014) discuss and confirm the ``growing mortality gradient by income'' and report a life expectancy gap of 13 years between those in the lowest versus highest average indexed monthly earning (AIME). It's worth noting that the correlation between (lower) income and (higher) mortality isn't only a U.S. phenomenon. 

Outside of the U.S. market, there is a similar discussion. Andersson, Lundborg and Vikstrom (2015) focus on Sweden, for example, where one wouldn't expect to observe such a mortality gradient. Milligan and Schirle (2018) provide recent evidence for the mortality income gradient in Canada. Nevertheless, the question motivating this paper is: how does this growing heterogeneity in mortality and longevity affect the value of annuitization?

Using numbers available in the late 20th century, Brown (2003) manufactures annuity prices and mortality tables based on race and education and concludes that ``complete annuitization is welfare enhancing {\em even} for those with higher than average mortality, provided administrative costs are sufficiently low.'' This result was echoed (and cited by) Diamond (2004) in his presidential address to the {\em American Economics Association}. He starts by noting that ``uniform annuitization would favor those with longer expected lives [such as] high earners relative to low earners'' but concludes that Brown (2003) ``shows much less diversity in the utility value of annuitization than previous comparisons."

And yet, the range of life expectancy between healthy and unhealthy in the Brown (2003) analysis (Table 1, to be specific) was 3.8 years at the age of 67. He assumed a conditional life expectancy of 81.0 years (lowest) for black males with less than a high school diploma vs 84.8 years (highest) for male Hispanics in the U.S. Contrast these differences with the more recent and granular numbers provided by Chetty et al. (2016), or even Goldman and Orszag (2014), where the gap in life expectancy between the highest and lowest income percentile (calibrated to the same age of 67) is between 10 to 15 years. It's unclear whether the uniformly positive {\em willingness to pay} for insurance values, i.e. {\em annuity equivalent wealth} values greater than one, can survive such a large gap in longevity expectations. The {\em annuity equivalent wealth} might be lower than {\em endowed} wealth  and the value of longevity risk pooling might be negative when retirees with low life expectancy are forcefully {\em pooled}, that is required to swim with individuals who are expected to live much longer. 

Can one say unequivocally that no matter how high Simon's mortality rate is, relative to Heather's, that he is willing to be pooled (with Heather) and benefits from annuitization? Surely there must be a point at which the answer is no, and he is willing to self-annuitize because the implicit loading, by having to subsidize Heather, is (too) high.

My point here isn't only to argue for an update or revision of possibly stale numbers to reflect the increasing heterogeneity of mortality, although that is certainly the motivation for the paper. Rather, my objective is also to focus attention on the mortality growth rate (and its inverse the longevity dispersion) as the driver of the value of annuitization. I illustrate this by utilizing a simple (closed-form) analytic expression for the annuity equivalent wealth and then calibrating to mortality rates by percentile, from the Chetty et al. (2016) data.

Conceptually the main argument of this paper can be summarized as follows. Although recent data indicates the heterogeneity of mortality is increasing and the gap in life expectancy is increasing, the same data confirm that individuals with higher mortality experience a higher volatility of longevity, as per the CLaM. This then serves to increase the value of longevity insurance. In other words, it isn't the shorter expected longevity (or higher mortality rate) that makes annuities appealing. Rather, it's the volatility of longevity that drives its value. More on this will be provided in the body of the paper.

To be clear, there are a number of other authors and papers that have focused attention and made the link between the {\em standard deviation} (SD) of lifespans and optimal lifecycle behavior. Most prominent in this category would be Edwards (2013), building on the work of Tuljapurkar and Edwards (2011), who document a 15 year standard deviation at the age of $x=10$. Edwards (2013) builds on the Yaari (1965) model and arrives at estimates for the increased longevity that a rational lifecycle consumer would demand in exchange for being exposed to a higher variance of life. Although much of Edwards (2011) is based on normally distributed lifespans, and I operate within a Gompertz framework, he shows that one additional year of standard deviation (in years) is ``worth'' about six months of life. Nevertheless, as far as the literature review is concerned, Edwards' is one of the few papers to focus economists' attention on the second (versus the first) moment of life and show {\em how exactly} it affects optimal behavior. From that perspective, this paper builds on the idea that the link between these two moments has an economic interpretation and implication.

\section{Matters of Life and Death}

\subsection{Mortality by Gender and Income}

Table (\#2) displays mortality rates for males and females in the U.S. as a function of various ages and income percentile. These represent realized mortality rates per 1,000 people during the period 2001 to 2014 and are based on the data collected by Chetty et al. (2016). The numbers provided are the actual ratio of observed deaths at a given age (say age 50) divided by the total number of people at that age (say 50). These rates aren't actuarial projections and are based on over 1.4 billion person-year observations and close to 6.7 million deaths. 

\begin{center}
\fbox{TABLE \#2a GOES HERE}
\end{center}

The methodology is described in the article by Chetty et al. (2016), and the entire dataset of mortality rates as a function of income percentile is available online. Their (lagged 2-year income) numbers are for ages 40 to 63, and one must employ forecasting procedures (a.k.a. Gompertz) to obtain values in later life, which is something I'll return to in a moment.

These mortality rates contain various insights or takeaways, some immediately obvious and intuitive and some (much) less so. Focus first on the middle row, with the so-called median mortality rates. At the age of 40, a total of 1.2 per 1,000 (median income) males died, whereas for (median income) females the rate was only 0.8 per 1,000 individuals. Stated differently, the one-year mortality rate for (median income) males at the age of 40 is 50\% higher than it is for females, which naturally leads to a lower life expectancy for (median income) males. Continuing along the same row, at the age of 50 the male mortality rate is now higher at 2.9 per 1,000 and for females it is 2.0, where I have dispensed with the phrases one-year and median income for the sake of brevity. At the age of 60, the rates are 7.3 (males) and 4.5 (females) respectively. This is simply the effect of aging, which affects both males and females. There is nothing surprising quite yet, but notice how the excess of male-to-female mortality shrinks from 150\% (=1.2/0.8) at the age of 40, to 145\% (=2.9/2.0) at the age of 50. This isn't quite a downward trend (at least in the table), since at the age of 60 the excess death is back to 162\% (=6.3/4.5).

Moving on to the (more interesting) rows, we now have the opportunity to measure the impact of income percentile on mortality rates. 

Notice that for a U.S. male in the lowest income percentile (south of the median), the mortality rate at the age of 50 is over four times higher at 12.5 deaths per 1,000, versus 2.9 at the median income. In stark contrast, a 50-year-old male at the highest income percentile (north of the median) experiences a mortality rate of only 1.1 per 1,000. This is less than half the median (income) rate. Stated differently, the range in mortality rates between the 1st percentile and 100th percentile is (12.5/1.1) or over eleven to one. To those who haven't seen such numbers before they might seem extreme, but they are by no means original. As noted by Deaton (2016) and quoted in the first paragraph of this paper, the link between mortality and income is well-established in the economics literature. 

Chetty et al. (2016), upon which these numbers are based, is simply one of the most recent and comprehensive documentations of the {\em mortality to income gradient.} In fact, Goldman and Orszag (2014) offer similar evidence and their data seem to indicate an even wider gap (i.e. greater than 15 years) in life expectancy based on income and wealth factors.

Digging a bit deeper, notice how the ratio of mortality rates between the lowest-income percentile (top of table) and the highest-income percentile (bottom of table) shrink or decline over time. For example, for males at the age of 40 the ratio of worst-to-best is 9.67 (=5.8/0.6), whereas at the age of 60 the ratio falls to 7.89 (=22.1/2.8). The same decline (in relative rates) is observed for females. At the age of 40 the ratio of worst to best is 14 (=4.2/0.3), but by the age of 60 it shrinks to a multiple of 5.8 (12.8/2.2). Stated differently, mortality rates appear to converge with age.

\subsection{Trends by Age: A Glimpse of Gompertz}

Table (\#2b) is constructed based on the numbers contained in Table (\#2a) and displays the annual rate at which one-year death rates themselves increase with age, as a function of income percentile. For example, the increase in one death-rates at the median income level (using a baseline age of 50) was 9.21\% per year for males and 8.64\% for females. The number in the right-most column in Table (\#2b) is the projected mortality rate at the age of 100 assuming the exponential growth continues at the same rate constant rate for 50 years, and is denoted by $\tilde{q}_{100}$ in the actuarial literature.

\begin{center}
\fbox{TABLE \#2b GOES HERE}
\end{center}

To be precise, this growth number is expressed in continuous time. It is computed by solving for $g$ in the relationship: $q_{63}=q_{50}e^{g13}$, or equivalently: $g=(\ln q_{63} / \ln q_{50})/13$ where $q_{63}$ denotes the mortality rate (for either males of females) at age: $x=63$ and $q_{50}$ is corresponding number at age: $x=50$. The upper age of 63 isn't arbitrary, but in fact is the highest age for which Chetty et al. (2016) report realized mortality rates as a function of (lagged 2-years) income.

The mortality {\em growth} rates of 9.21\% for males and 8.64\% for females (or approximately 9\% on a unisex basis) at the median income level aren't restricted to the age range of 50 to 63 and are not an artifact of this particular dataset. The (approximate) 9\% rate growth in mortality is observed in most human species from the age of 35 to the age of 95. It is known as the Gompertz (1825) law of mortality, named after Benjamin Gompertz.

Back to the topic of mortality inequality, notice though how the growth (rate) of mortality (rates) for individuals at the lowest income percentile is only 5.63\% for males and 4.81\% for females, which is close to half of the corresponding rate at the median income level. Perhaps this can be interpreted as some modicum of good news for the less economically fortunate. Their mortality rates don't grow or increase as fast. Of course, they have started off (at the age of 50) from a much higher base. In contrast, those fortunate to live in the highest income percentile experience a 10\% growth in mortality as they age. Stated differently, they {\em age faster} than the median person in the population and {\em much} faster than those at the lowest income percentile. In fact, the difference between males (10\%) and females (9.97\%) is almost negligible. Notice how by age $x=100$ the mortality rates are much closer to each other. I'll get back to this.

The disparity in mortality {\em growth} rates between high and low-income individuals leads to a corresponding gap in the {\em dispersion} of the remaining lifetime random variable: $T_x$. I'll define this variable formally and precisely, but there is in fact an inverse mathematical relationship between the mortality growth rate and the standard deviation: $SD[T_x]$. The lower growth rate is synonymous with an increase in coefficient of variation of longevity (CoVoL) which is defined as the ratio of the standard deviation of remaining lifetime $SD[T_x]$ to the mean remaining lifetime $E[T_x]$. The demand for longevity insurance is relatively higher -- and {\em ceteris paribus} they are willing to pay more for insurance -- compared to those at the lowest income percentile. A formal discussion is presented in Section (\#4.)

\subsection{Compensation Law of Mortality (CLaM)}

Figure (\#2) is a visualization of the so-called compensation law of mortality, which explains (and is consistent with) the data displayed in tables (\#2a) and (\#2b).

\begin{center}
\fbox{FIGURE \#2 GOES HERE}
\end{center}

The convergence of the mortality rate curves (or regression lines when plotted against $\ln q_x$), was to this author's knowledge first identified by L.A. Gavrilov and N.S. Gavrilova and fully explained in the book by Gavrilov and Gavrilova (1991, page \#148) using what they call a reliability theory of aging\footnote{The negative relationship (between $c_0$ and $c_1$) is occasionally referred to as the Strehler-Mildvan correlation but isn't quite the nature of the link described above. See also the interesting and related paper by Marmot and Shipley (1996), which suggests that mortality rates may not converge as fast at advanced ages.}. I follow Gavrilov and Gavrilova (1991, 2001) and assume Gompertz to (more) advanced ages than assumed by Chetty et al (2016), although whether or not mortality plateaus at age 100 or perhaps 105 isn't quite pertinent to the discounted value of a life annuity at age 65.

Without drifting too far from the script of this particular article, although the {\em Compensation Law of Mortality} and the negative relationship between (i.) middle-age death rates and (ii.) growth rates in any sub-group of species is known to biologists, it remains controversial because it's difficult to reconcile with popular theories of aging and longevity. Nonetheless, a framework that can explain this compensation law is the so-called reliability theory of aging, which also explains why organisms prefer to die according to the Gompertz law of mortality. I refer readers to Gavrilov and Gavrilova (2001) for more on the underlying {\em biology of lifespan} and return to (the safety of) economic risk and return.

\subsection{Review of the Gompertz Law of Mortality}

To analyze the impact of the longevity variability on the demand for insurance one requires a parsimonious model for mortality rates over the lifecycle, and not only a few discrete age data points. Recall from the earlier subsection that mortality rates $q_x$, increase exponentially in age $x$, and $\ln q_x$ increases linearly in $x$. This suggests that in continuous time, a suitable model for the log {\em hazard} rate is:
\begin{equation}
\lambda_{x+t} \; = \;  h_xe^{gt} \; = \; \frac{1}{b} e^{\frac{x+t-m}{b}} \; = \; \big( \frac{e^{(x-m)/b}}{b} \big)e^{t/b}
\label{G1}
\end{equation}
This is the formal Gompertz law of mortality mentioned earlier. The $(m,b)$ parameterization is common in the actuarial finance literature. The simpler $(h,g)$ formulation is more common in demographics, statistics and economics. I will use both of them interchangeably, depending on context and need.  

Under the $(m,b)$ formulation, two free (or governing) parameters have a more intuitive probabilistic interpretation. They are labeled the {\em modal} value $m$, and the {\em dispersion} parameter $b$, both in units of years. Different groups within a species or population have different $(m,b)$ values, in particular those with lower incomes will be characterized by low values of $m$ and high values of $b$. Or, stated in terms of $(h,g)$, at any given age, the hazard rate $h$ for low income groups is higher, but the growth rate $g$ is lower.

The intuition behind the $(m,b)$ perspective -- and why they are labeled modal and dispersion -- only becomes evident once the remaining lifetime random variable: $T_x$, is defined via the hazard rate with a corresponding density function. Formally:
\begin{equation}
\Pr[T_x \geq t] := p(x,t,m,b) = e^{-\int_0^t \, \lambda_{x+s}ds} = \;  \exp \{e^{(x-m)/b} (1-e^{t/b}) \},  
\label{G2}
\end{equation}
where going from the third to the fourth term in equation (\ref{G2}) follows from the definition of the hazard rate in equation (\ref{G1}). The expected value of the remaining lifetime random variable: $E[T_x]$, as well as any higher moment can be computed via the cumulative distribution function (CDF) induced by: $F(x,t,m,b):=1-\Pr[T_x \geq t]$, or more commonly using the probability density function (PDF) defined by: $f(x,t,m,b)=-\frac{\partial}{\partial t} p(x,t,m,b)$. One can re-write the density function in terms of hazard rate and growth rate, or $(h,g)$ as well.

The {\em coefficient of variation of longevity} (CoVoL) measures the relative uncertainty of the individual's longevity or retirement horizon. In this paper I will occasionally refer to this ratio as the {\em volatility of longevity}, since it is frequently used in this context within the financial services industry as a common measure of longevity risk. Regardless of its exact name, I'll denote it by the symbol $\varphi_x$ and make sure to index by current age $x$, because it can obviously change over the lifecycle. It is defined equal to the standard deviation of the ratio of: $T_x$ to its expectation: $E[T_x]$. Formally it can be computed as follows:
\begin{equation}
\varphi_x \; = SD ( \frac{T_x}{E[T_x]} ) \; = \frac{SD[T_x]}{E[T_x]} \; =
\; \frac{\sqrt{\int_0^{\infty} \, t^2 \, f(t)dt - (\int_0^{\infty} \, t \, f(t)dt)^2}}{\int_0^{\infty} \, t \, f(t)dt}.
\label{ivol}
\end{equation}
Note that the abbreviated: $f(t)$ is shorthand notation for the full probability density function (PDF) under a Gompertz law of mortality. Figure (\#1), which I mentioned earlier in the introduction when comparing Heather to Simon, is essentially a plot of $f(t)$ under two sets of Gompertz $(m,b)$ values. I'll present a number of examples and explicitly compute values of $E[T_x]$, $SD[T_x]$ and $\varphi_x,$ in the next sub-section. 

Before I get to numbers, I would like to help readers develop some intuition for the $\varphi_x$ function by examining its properties under the simplest of (non-Gompertzian) mortality laws, namely when: $\lambda_x=\lambda$ and is constant at all ages. Humans age over time and their hazard rate increase, but there are actually a few species that never age. In our language, their hazard rate remains constant regardless of how old they are. They die (obviously) but the rate at which this occurs never changes. In fact, a constant hazard rate is associated with an exponential remaining lifetime and $\Pr[T_x \geq t]=e^{-\lambda t}$. The probability of survival declines with $t$, but the rate doesn't change. Under this non-Gompertz law, the expected value: $E[T_x]=\lambda^{-1}$ and the standard deviation is (also) $SD[T_x]=\lambda^{-1}$; both are well-known properties of exponential lifetimes. 

Back to the volatility; according to the general definition presented in equation (\ref{ivol}), but applied to the exponentially distributed lifetimes; $\varphi=1$, at all ages and under all parameter values. In other words, the CoVoL is invariant to $\lambda$. In the common parlance, longevity risk is always 100\% when the mortality rate is constant. Indeed, this is more than a nice coincidence and is in fact the reason I chose to focus on the CoVoL. Back to the Gompertz law of mortality, the CoVoL is (i.) always less than 100\%, (ii.) increases with age over the lifecycle, and (iii.) higher for those with higher mortality rates at the same chronological age, e.g. Simon versus Heather. So, to be crystal clear, both the SD of: $T_x$ and the CoVoL of $T_x$ are higher for Simon versus Heather.

\subsection{CoVoL numbers and intuition}

Table (\#3) offers a range of numerical values for the function: $\varphi_x$, at the age of $x$, under an assortment of parameter values for $m$ (the mode of the Gompertz distribution) and $b$ (the dispersion coefficient.) Note that by fixing $(m,b)$ the hazard rate at any age $x$, is simply: $\lambda_x=(1/b)e^{(x-m)/b}$, as per the definition of the Gompertz mortality hazard rate.

For example, the values in the first row are computed assuming that $m=98$ and the mortality growth rate is $1/b=11.5\%$ per year, which implies that the dispersion coefficient is: $b=6.696$ years. For reasons that should by now be obvious, I refer to this combination of $(m,b)$ as rich (i.e. highest income percentile), and the life expectancy at birth: $E[T_0]=92.98$ years. The standard deviation of the lifetime random variable at birth is: $SD[T_0]=11.15$ years, which is higher than the dispersion coefficient $b=8.696$ by slightly more than two years. At birth, the standard deviation: $SD[T_0] = b \, \pi/\sqrt{6}$, which is 28\% greater than $b$.

To be clear, computing the mean and standard deviation at birth under a Gompertz assumption for the evolution of the hazard rate $\lambda_x, x=0..\omega,$ is somewhat artificial (or disingenuous). After all, even in the most developed countries mortality rates are relatively high in the first few years of life, reaching a minimum around the age of 10 and only starting to ``behave'' in a Gompertz-like manner between the age of: $x=30$ to $x=95$. I'm not saying all mortality tables at all ages are Gompertz. Rather, the point here in table (\#3) is to provide some intuition for the moments of the Gompertz random variable as opposed to accurately modeling the earliest ages in the lifecycle. Needless to say, infants and children aren't purchasing life annuities or pooling longevity risk.

Moving along the first row of the table, the CoVoL is a mere: $\varphi_0(98,8.686)=12\%$ at birth, then increases over the lifecycle to reach: $\varphi_{65}(98,8.686)=33.7\%$ at the age of $x=65$. Note -- for the sake of replicability -- that although $\varphi_0$ can be derived analytically the CoVoL numbers at $x>>0$ are computed numerically. The denominator of $\varphi_x$, that is the mean of the remaining lifetime, is available in closed form (see Appendix A.3 with $r=0$), while the numerator requires a numerical procedure. In sum, the rich woman's CoVoL at the age of retirement age of $x=65$ is $33.7\%$. Again, the $m=98$ and $b=8.696$ parameters are representative of the highest income percentiles in the Chetty et al. (2016) data. This 65-year-old has a mortality hazard rate of $\lambda_{65}=0.2586\%$ at the age of $x=65$.

Now, as I move down the rows within the first panel and artificially reduce the mortality growth rate from 11.5\% to 5.5\%, while holding the modal value of life fixed at 98 years, the CoVoL values increase. To be clear, holding $m$ fixed and reducing $g:=1/b$ induces a corresponding increase in $\lambda_x$ as per the 6th column in the table. Although these parameter combinations don't correspond to any particular income percentile the point here is to develop intuition for the link between age, parameters $(m,b)$, hazard rates $\lambda_x$ and CoVoL. 

Notice how reducing the mortality growth rate (remember this is: $g=1/b$), that is increasing $b$, uniformly reduces the life expectancy at birth $E[T_0]$ and increases the standard deviation $SD[T_0]$ at birth. The numerator goes up, the denominator goes down and so the CoVoL increases with higher values of $b$. 

The situation becomes less obvious or clear-cut at the age of 65. Notice how the mean $E[T_{65}]$ starts-off at 28.82 years (when $b=8.696$) and then declines as one would expect at higher values of $b$, but then flattens-out at 28.59 years and starts to increase. It seems the increased dispersion $b$ then serves to increase the mean value, as often happens with highly skewed distributions. This is also driven by the fact the initial hazard rate at age 65 is higher. All in all, the standard deviation $SD[T_{65}]$ does in fact increase monotonically with $b$ and the net effect is that the coefficient of variation does in fact increases in $b$.

Moving from rich to poor, in the bottom panel of table (\#3) in which the modal value of life is $m=78$ years, the corresponding life expectancy values are (much) lower at ages. The hazard rates are higher as well. At the very bottom row, where $b=18.182$ years (i.e. the slowest aging), the life expectancy at retirement age $x=65$ is now 17 years (vs. 28 years). The poor man's CoVoL is 61.9\%, which is almost double the top row, compared to the rich woman's 33.7\%.

The artificial numbers in table (\#3) and the underlying intuition are quite important for understanding the comparative statics in the next section. Notice how the CoVoL is subtly affected by the interplay between age $x$, modal value $m$ and dispersion value $b$. As $x \rightarrow m$ the CoVoL increases purely as a result of the underlying Gompertz law, even when $b$ is unchanged. In other words, aging increases the relative riskiness of your remaining lifetime. From an economic point of view -- as I will show in the next section -- the demand for longevity insurance and risk pooling increases in age, a.k.a. with higher mortality rates. Just as importantly, fixing both $x$ and $m$, and (only) increasing the value of $b$ also increases CoVoL. At any given age $x$, your CoVoL is higher with higher $m$ and/or with higher $b$. Figure (\#3) displays CoVoL numbers over the entire lifecycle for two sets of $(m,b)$ parameters. Notice how it plateaus at higher ages and never exceeds 100\%. One can think of it as the expected value of the Gompertz remaining lifetime: $E[T_x]$ converging to $SD[T_x]$, as $x \rightarrow \omega$. In probabilistic terms, this is how to think about it. At very advanced ages the amount of time you {\em will actually} live is very close to what you {expect to} live. See Figure (\#3).

\begin{center}
\fbox{FIGURE \#3 GOES HERE}
\end{center}

For clarity, the variability is unrelated to notions of stochastic mortality, Lee and Carter (1992) models, or aggregate changes in population $q_{65}$ values over time. I am not using volatility to forecast mortality in (say) 50 years and the models in this paper are entirely deterministic. That said, if one graduates from a deterministic model of hazard rates $\lambda_x$, and postulates a stochastic model: $\tilde{\lambda}_x$, the $\varphi_x$ metric would still be defined as the ratio of moments and should converge to a value of one at very advanced ages. 
\section{Annuity Equivalent Wealth and Willingness to Pay}

We now arrive at the main tool I will use to measure whether Simon with the higher CoVoL, benefits -- in utility terms -- from being forced to purchase loaded pension annuities.

\subsection{Pricing Annuities}

I use the symbol {\tt a}$(.)$ to denote the market price of a life annuity at age $x$. In words, paying: {\tt a}$(.)$ to an insurance company or pension fund, obligates the issuer to return $\$1$ of income per year (or $\$1dt$ in continuous time) for the life of the buyer, a.k.a. retiree or annuitant. The items inside the bracket $(.)$ are the conditional factors, which could be age, gender, health, etc. For example, a retiree might pay: {\tt a}$(65)\,=\$20$ to purchase a life annuity providing $\$1$ per year for life starting at the age of: $x=65$, but the price might be: {\tt a}$(75)\,=\$14$ if purchased at age: $x=70$, etc. Both are completely arbitrary numbers, although throughout the paper for the sake of this discussion I'll assume life annuities scale and the price of $\$100,000$ of annual income is exactly $\$100000${\tt a}$(.)$. There are no bulk (economies of scale) discounts or (adverse selection-induced) penalties. Moreover, the only loading or frictions will come (implicitly) from being pooled with healthier and longer lived individuals, as opposed to profits or other institutional features. Now, in the above example market prices are differentiated by age $x$. In practice (most) companies and issuers differentiate by gender, while some go even further and {\em underwrite} annuities, that is price based on the health status of the annuitant. When needed, I will augment notation to include biological characteristics, namely the two Gompertz parameters $(h_x,g)$ at the relevant to age $x$.

Thus, {\tt a}$(h_x,g)$ denotes the market price of a $\$1$ per year life annuity, purchased at age: $x$ when the hazard rate is $h=h_x$, for an individual whose remaining lifetime random variable is modeled in Gompertz $(h_x,g)$ space. These were all explained in prior Section \#3. The point I make here is that these two bio-demographic characteristics are easily observable and can (legally) be used for underwriting -- or at least for theoretical {\em valuation} purposes.

As far as finance and markets are concerned, interest rates (obviously) impact the price of pension and life annuities, so in the event I must draw attention to the underlying pricing rate: $r$, assuming it is constant, I will append a third parameter to the very beginning of the expression and write the annuity factor as: {\tt a}$(r,h,g)$ for completeness. Notice that absence of any age $(x)$ in the expression, since this is already contained and included within the hazard rate $(h_x)$. Occasionally the expression {\tt a}$(r,x,m,b)$ will make an appearance when I want to draw specific attention to the impact of a modal value $(m)$ or global dispersion parameter $(b)$, on the annuity valuation factor at an explicit age denoted by $(x)$.

Notice that up to now I have only mentioned market {\em prices} as opposed to say theoretical model or economic {\em values}, which naturally might differ from each other. To link these two numbers, I refer to what actuaries might call the fundamental law of pricing mortality-contingent claims, or what financial economists might simply call {\em No Arbitrage} valuation. Either way, the market price: {\tt a}$(x)$ under a Gompertz law of mortality \underline{should} be equal to the following value:
\begin{equation}
{\tt a}(r,h_x,g) \; := \int_{0}^{\omega - x} \, e^{-rt} \, p(t,h_x,g) dt,
\label{AF1}
\end{equation}
where $\omega$ denotes the last possible age to which people are assumed to live and $p(t,h_x,g)$ is the conditional (at age $x$ and mortality rate $h_x$) survival probability. The underlying economic assumption is that if a large enough group of known $(h_x,g)$-types are pooled together they will -- by the law of large numbers -- eliminate any idiosyncratic mortality risk and the valuation is easily conducted by discounted cash-flow expectations. Another embedded assumption in equation (\ref{AF1}) is that the interest rate is a constant $r$, although that really isn't critical. By assuming a constant rate $r$, equation (\ref{AF1}) can be solved analytically using the Gamma function. See the technical appendix for more on this.  Equation (\ref{AF1}) is often used discretely, for example in Poterba, Venti and Wise (2011). From this point onward I will refrain from using market {\em price} or {\em value} and refer to: {\tt a}$(.)$ as the annuity {\em factor.}

Before I proceed to the economics of the matter, it's important to focus attention on the sign of the partial derivatives of: {\tt a}$(r,h_x,g)$, with respect to the three explicitly listed arguments. First, the annuity factor declines with increasing age and hazard rate $h_x$. Intuitively (and unlike a perpetuity) the cost of a constant $\$1$ of lifetime income declines as you get older (and closer to death). Likewise, the factor declines at higher valuation rates $r$, after all it's just a present value. Just as importantly, at any conditional age $x$ and hazard rate $h_x$, the annuity factor declines as the growth rate $g$ is increased, which is synonymous with individuals who have a lower remaining life horizon. These insights don't require much calculus and are discussed at greater length in the technical appendix.

Back to the economics of the matter. Let $(\hat{h}_x,\hat{g})$ denote the Gompertz parameters that best fit population (group) mortality while $(h^i_x,g^i)$ denotes the parameters that best fit individual (sub-group) mortality. In particular, using ideas introduced in section \#3, I let: $(h^{1}_x,g^{1})$ denote the Gompertz parameters of individuals (i.e. sub-group) in the lowest income percentile, whereas $(h^{100}_x,g^{100})$ denotes the Gompertz parameters of individuals in the highest income percentile. Therefore, the population modal and dispersion parameters would be the median values: $\hat{h}=h^{50}_x,$ and $\hat{g}_x=g^{50}$, respectively (albeit with a bit of hand waving).\footnote{Note. The weighted average of Gompertz variables isn't Gompertz. Second, even if I select the best fitting average line, the $h_x$ and $g$ values will {\em not} be linear averages. So, this is an approximation, but precisely what Chetty et al. (2016) assumed as well.} As mentioned earlier, on the occasion that I want to draw attention to population averages for $(m^i)$ and $(b^i)$ parameters, I'll use the obvious: $(\hat{m})$ and $(\hat{b})$.

\subsection{Measuring Utility}

Let: $U^{**}(w)$ denote the value function (maximal utility) of the individual who annuitizes their wealth $w$, and $U^{*}(w)$ the value function of the individual who decides to self-insure (i.e. not own any annuities at all) and instead decides to fund retirement with a systematic withdrawal program, then:
\begin{equation}
U^{**}(w) \; \geq U^{*}(w).
\end{equation}
This is the well-known result in annuity economics, noted and cited in section \#2. There exists a constant $\delta \geq 0$ such that:
\begin{equation}
U^{**}(w) \; = U^{*}(w(1+\delta)).
\end{equation}
A retiree who doesn't annuitize would require the $\delta$ percent increase in their wealth $w$ to induce the same level of utility as someone who does annuitize. Given that we are operating with constant relative risk aversion (CRRA) utility and no pre-existing annuity income, I will set $w=1$ and refer to the {\em annuity equivalent wealth} by: $(1+\delta)$, and the value of longevity risk pooling by $\delta$. To close the loop on all these (utility based) definitions, note that if my subjective value of risk pooling is: $\delta=25\%$, and the AEW is $\$1.25$, then someone with $w=\$125$ of initial wealth would be willing to pay $\$25$ or \$100$\delta / (1+\delta)$ to have access to the annuity. The {\em willingness to pay} is then $\delta/(1+\delta)$. 

Whether it be AEW, WtP or simply {\em the value of pooling} $\delta$, all essentially measure the same thing and will be used interchangeably in the paper, unless the numbers themselves are important. Either way, the $\delta$ is a function of the individual Gompertz parameters $(h^i_x,g^i)$, the market pricing parameters $(\hat{h}_x,\hat{g})$, and the utility-based preferences involving risk $\gamma$ and discounting $\rho=r$.

Moving on, assume an individual (denoted by $i$) and their force of mortality are individually Gompertz parameters and the population is also (approximated as) Gompertz, upon which annuity factors are based. The individual's AEW can be expressed as:
\begin{equation}
1+ \delta^i_x \; = \frac{{\tt a}(r,x,m^i,b^i)^{1/(1-\gamma)} \, {\tt a}(r,x,\hat{m},\hat{b})^{-1} }{{\tt a}(r,x-b^i \ln[\gamma],m^i,b^i)^{\gamma/(1-\gamma)}}  \; = \; 
\frac{{\tt a}(r,h^i_x,g^i,)^{1/(1-\gamma)} \, {\tt a}(r,\hat{h}_x,\hat{g})^{-1} }{{\tt a}(r, h^i_x/\gamma,g^i)^{\gamma/(1-\gamma)}}  
\label{delta1}
\end{equation}
where $\gamma$, denotes the coefficient of relative risk aversion within CRRA utility, and assuming the subjective discount rate is (also) equal to $r$. 

I have deliberately expressed the annuity factor using both $(m,b)$ and $(h_x,g)$ formulations, mainly so that $\delta$ can be (easily) computed when either set of parameters are available. Note that when annuity prices are fair (i.e. pooling with equal risks) the value of $h^i_x=\hat{h}_x$, and as well the value of $g_i=\hat{g}$. So, the expression in equation (\ref{delta1}) can be simplified further, which I will do in a moment.

The value of longevity risk pooling $\delta$ is an increasing function of the coefficient of variation of longevity (CoVoL) for the individual. The precise mechanism by which this operates is via the higher mortality hazard rate $h_x$. To be clear, even if $g$ is held constant, a higher value of $h_x$ induces a higher (CoVoL and) value of $\delta$. This, again, is when annuities are fairly priced so that $h_x=\hat{h}_x$ and $g=\hat{g}$ in equation (\ref{delta1}). Supporting proofs and comparative statics are presented in the technical appendix. I also refer to Milevsky and Huang (2018) for additional supporting material. Note that my interest in this paper is to {\em use} the expression for annuity equivalent wealth, a.k.a. ($1+\delta$).

Finally, in the homogenous case everyone in the sub-group experiences the exact same Gompertz force of mortality, that is $h_x=\hat{h}_x$ and $g=\hat{g}$. The {\em annuity equivalent wealth} is:
\begin{equation}
1+ \delta_x \; = \;  \left( \frac{ {\tt a}(r,x,m,b)}{ {\tt a}(r,x-\hat{b}\ln\gamma,m,b)} \right)^{ \frac{\gamma}{1-\gamma}} \; = \;
\left( \frac{ {\tt a}(r,h_x,g)}{ {\tt a}(r,h_x/\gamma,g)} \right)^{ \frac{\gamma}{1-\gamma}}.
\label{delta2}
\end{equation}
It is a simple function of the ratio of two actuarial annuity factors and easily computed in either discrete or continuous form.

In particular, notice that the respective annuity factors in equation (\ref{delta2}) are computed at two distinct ages (or hazard rates.) The numerator is computed at the biological age: $x$ (or hazard rate $h_x$), and the denominator is computed at a modified (risk-adjusted) age: $x-\hat{b}\ln\gamma$ (or hazard rate $h_x/\gamma$.) The modified age in the denominator's factor is under (younger than) $x$ whenever $\gamma >1$. The lower modified age increases the annuity factor and the {\em AEW}. Nevertheless, even when $\gamma<1$, the value of $\delta_x>0$ provided $\gamma>0$. Here are some numerical examples and intuition for the {\em annuity equivalent wealth}.

\subsection{A Simple Numerical Example: Two Hypothetical Groups}

Assume the remaining life expectancy at the age of 65 of a hypothetically constructed {\em Group A} is: $11.95$ years and the underlying Gompertz mortality parameters are: $m=75.02$ and $b=11.87$. The CoVoL at age 65 is $60.4\%$, which recall, is the standard deviation of remaining lifetime: $SD[T_{65}]$ scaled by mean remaining lifetime: $E[T_{65}]$. Now, assume the remaining life expectancy for a hypothetical {\em Group B} is: $E[T_{65}]=23.64$ years and Gompertz parameters are: $m=91.72$ and $b=12.87$, for a coefficient of variation of longevity (CoVoL) of: $47.4\%$. 

As this is only a numerical example, I'll assume that the (subjective discount rate and) valuation rate are: $r=3\%$, and the coefficient of relative risk aversion is: $\gamma=3$ in the CRRA utility. These values are not unreasonable for the valuation of subjective utility and AEW.

Under these parameter values a representative Group A (65 year old, retiree) would value the longevity insurance at: $\delta = 89.32\%$ if they could pool this risk with other similar Group A members, by acquiring fairly priced life annuities based on their own population: $m=75.02$ and $b=11.87$ parameters. This $89.32\%$ number is based on equation (\ref{delta2}), first by computing {\tt a}$(0.03, 65,75.02,11.87)=9.493$, dividing by {\tt a}$(0.03, 51.96,75.02,11.87)=14.528$, raising to the power of $(-3/2)$ and then subtracting one to express as a percentage. In particular, notice the age setback in the denominator from: $x=65$ to the modified age of $51.96=(65-(11.87)\ln3)$, which I labeled a risk-adjusted age. This AEW is in line with the (generally) large benefits from annuitization reported in (many) other studies over the last 30 years, noted in the literature review of section \#2. For example, Brown (2003) computes $\delta$ (numerically) over a range of coefficient of relative risk aversion (CRRA) values, $\gamma=1..5$. His numbers range from 36\% to 90\%, depending on demographic factors.

Now, let's examine the representative Group B retiree, assuming the same $r=3\%$ valuation rate and $\gamma=3$, coefficient of risk aversion. If they are pooled with homogeneous risks from Group B, their $\delta$ value at the age of $x=65$ is lower than their Group A counterparts. Intuitively, their CoVoL is also lower at age 65. In particular, using the $m=91.72,b=12.87$ values, equation (\ref{delta2}) leads to a value of $\delta=48.39\%$ which is lower than what a Group A member would be willing to pay. This is despite the obvious (but not quite relevant) fact that the Group B retiree is expected to live 24 years versus the 12 years for the Group A member. Indeed, what drives the $\delta$ for (fair) longevity insurance is the volatility of longevity (via the $m$ and $b$) and not the demographic fact that the Group B retirees live longer. 

Moving on to a larger heterogenous pool, imagine Group A and Group B, both at chronological age 65, are mixed together in equal amounts in a large pool. To keep the system fair, the pension annuities are priced based on the mixed population mortality. With a bit of hand waving, assume the resulting Gompertz parameters upon which the group annuities are priced are: $m=85.45$ and $b=12.41$. Intuitively, the Group A member is presented with a relatively worse annuity price and the Group B member is getting a relatively better price. 

This assumes both groups are forced to purchase annuities at the same price in a compulsory system. The {\em AEW} is based on the group population (for market pricing) and individual (for lifecycle utility) mortality. See equation (\ref{delta1}). The group annuity factor -- or price they both pay -- is now {\tt a}$(r=3\%, 65,\hat{m}=85.45,\hat{b}=12.41)=13.583$, which isn't actuarially fair to either of them. It's advantageous to the Group B member (who would have had to pay $15.97$) and disadvantageous to the Group A member (who would only have to pay $9.493$). 

Here is the main point. The Group B member now experiences: $\delta=74.48$, which is higher than the prior (homogenous case of): $\delta=48.39\%$.  In contrast, the Group A member is faced with loaded prices (13.583 vs. 9.493) and they are only valuing the annuity at: $\delta=32.32\%$. In fact, if the Group A member had a life expectancy that was (even) lower, but still paid the same group price, it's conceivable the $\delta$ in equation (\ref{delta1}) might actually be negative. Under those conditions Group A (fish) would not be willing to swim in a pool with healthy Group B (sharks). They would rather take their chances and self-insure longevity. 

\begin{center}
\fbox{FIGURE \#4 GOES HERE}
\end{center}

Figure (\#4) illustrates this relationship graphically over a spectrum of possible $m$ values. On the left are individuals (think Group A) with low life expectancy (proxied by $m$) values and a correspondingly higher volatility of longevity at retirement. To the right are individuals (think Group B) with higher life expectancy and lower volatility of longevity. The Gompertz dispersion of longevity -- in contrast to the {\em volatility} of longevity -- for the purposes of computing equation (\ref{delta1}), is held constant at: $1/g=b=12$ years. The upper range is based on $\gamma=5$, that is higher levels of risk aversion, whereas the lower range is for $\gamma=1$, albeit with slight modifications in equation (\ref{delta1}) to account for logarithmic utility.

Notice that as $m$ increases, the value of longevity risk pooling and willingness to pay declines -- when the pooling is homogenous. A Group B retiree (paying fair prices) experiences a $\delta$ that is lower than a Group A (paying fair prices); but when they are mixed together and both pay the same group price, the curve is reversed. Intrinsically the Group A values longevity insurance more, due to his/her higher volatility of longevity, but the positive loading reduces its appeal. In contrast, the Group B member who -- shall we say loosely -- was somewhat ambivalent about the benefit of longevity risk pooling, is now willing to pay more for a relatively cheap life annuity. 

It might seem odd to talk about a change in the subjective willingness to pay for something based only on its market price. After all, my willingness to pay for apples or oranges should not depend on price. Indeed, that's the point of the metric, to be invariant to market prices. Rather, the proper way to think about this as the willingness to pay to have some oranges (i.e. Group A) mixed in with your apples (i.e. Group B) when annuitizing. The healthy and wealthy Group B member (shark) is willing to pay more for the mixed bag. 

With that intuition out of the way, we can back to our main empirical question. What happens in the U.S. across different income percentiles? Does the lowest income percentiles (i.e. Simon) experience a large enough volatility of longevity to over-come the implicit loading induced by having to subsidize the annuities of the highest income (i.e. Heather)?

\section{AEW by Income Percentile in the U.S.}

I now use the main equation (\ref{delta1}) for the value of longevity risk pooling, a.k.a. the annuity equivalent wealth, with plausible values for the coefficient of relative risk aversion, $\gamma=1..5$, and actual values for the Gompertz parameters $(m_i,b_i)$ as a function of income percentile. 

\begin{center}
\fbox{FIGURE \#5 GOES HERE}
\end{center}

As per the {\em compensation law of mortality} which was explained and introduced earlier, Figure (\#5) displays the relationship between $(h)$ and $(g)$ as a function of income percentile, based on the earlier-mentioned data contained in Chetty et al. (2016)\footnote{As far as measuring risk aversion is concerned, I'm aware of the ongoing debate and well known problems in calibrating $\gamma$ and refer the interested reader to recent work by O'Donoghue and Somerville (2018), or Schildberg-H\"{o}risch (2018) and rely on Andersen (2008), for example, for the justification in using such values. Likewise, while making comparison across different percentiles, I'll assume that $\gamma$ remains constant and doesn't depend on the particular choice of $(m_i,b_i)$ or $(h^i,g^i)$ values. In fact, an argument could be made that individuals with higher mortality might actually have lower levels or risk aversion. See Cohen and Einav (2007) for a possible link between risk exposure and demand for insurance.}. Note that the same negative relationship between initial mortality rate and the mortality growth rate is observed with Canadian data, as per Milligan and Schirle (2018). It is illustrated is Figure \#6. So, this is not a spurious relationship that only holds in the U.S.

\begin{center}
\fbox{FIGURE \#6 GOES HERE}
\end{center}

Table (\#4) displays results using the mid-point of $\gamma=3$, with additional (tables of values) available upon request. First, I estimate the relevant Gompertz $m^i$ and $b^i$ (or $h^i$ and $g^i$) values, then I compute the remaining lifetime expected value: $E[T_x]$ and standard deviation: $SD[T_x]$. With those numbers in hand, I can show the {\em individual volatility of longevity} and finally the two AEW values. Notice how both the Gompertz $b$ value, which measures the random variable's intrinsic dispersion, and the volatility of longevity decline at higher income percentiles. Again, the CoVoL is lower due to the decline in $b$ and the intrinsic increase of CoVoL as age $x$ edges closer to the value $m$. 

\begin{center}
\fbox{TABLE \#4 GOES HERE}
\end{center}

The last two columns in table (\#4) are computed using equation (7) and (8), for individuals and groups respectively. To be clear, the column labeled AEW for Individual, assumes that at any percentile, these 65-year-olds are pooled with individuals who share their same $m_i,b_i$ and are identical risk types. They pay fair actuarial prices for their annuity. Moving down the panels, higher-income percentiles are associated with (lower CoVoL) and lower values of annuity equivalent wealth under individual pricing. The intuition for this result was presented in section (\#4.1) and displayed in figure (\#4). Notice how much more annuities are worth to the poor vs. rich, when they are fairly priced.

Moving on to mandatory annuities pricing, the right-most column, is based on loaded pricing which is obviously disadvantageous for some (poor) and beneficial for others (rich.) As noted many times in the paper, if you are healthier than some people in the annuity pricing pool -- and you aren't paying the fair rate for your risk -- group pricing offers a higher annuity equivalent worth. The group value of AEW {\em increases} with income percentile for males, because they get better relative discounts. This pattern (and one that was illustrated hypothetically in figure \#4) is not observed for females. The reason for this discrepancy, or lack of uniformity, is that even though the annuity is {\em loaded} and relatively unappealing to the lowest income group, the loading isn't high enough to generate a lower annuity equivalent wealth for the lowest income compared to the higher income. In some sense the uniformly increasing pattern of the (Group) $\delta$ value for males is the surprising result. Either way, the important point here is that {\bf all} values of $\delta>0$, for the $\gamma=3$ case. They all benefit (as far as utility is concerned) from swimming together.

Again, the key (policy) takeaway from this table is that {\em even} at the lowest income percentile, where the Gompertz hazard rate is higher and life expectancy at retirement is a mere decade or so, the value of $\delta$ is positive {\em even} at the group pricing rate. Of course, this assumes that prices are purely based on group mortality, that is determined by the 50-percentile parameters. If there are additional loadings or costs added-on, the $\delta$ might become negative. I should note that in my extensive analysis (not all displayed) at low income percentiles, large amounts of pre-existing pension annuity income and lower levels of risk aversion $\gamma$, the $\delta$ value is barely positive.

\section{Summary and Conclusion}

In a mandatory pension system participants with shorter lifetimes {\em ex ante} subsidize those expected to live longer. Moreover, since individuals with higher incomes tend to have lower mortality, the poor end-up subsidizing the rich. To quote Brown (2003), ``When measured on a financial basis, these transfers can be quite large, and often away from economically disadvantaged groups and towards groups that are better off financially." This uncomfortable fact is well established in the literature and occasionally touted as a (social) justification for transitioning to Defined Contribution (DC) schemes. And yet, Brown (2003) goes on to write that ``the insurance value of annuitization is sufficiently large that relative to a world with no annuities, all groups can be made better off through mandatory annuitization." 

The question motivating this paper is: At what point does the gap in longevity expectations become such that the value of annuitization is actually negative for the ones who are expected to live the least? This is an empirical question and quite relevant to the growing disparity in U.S. mortality as a function of income. Against this background, this paper focuses attention on the heterogeneity of the second moment of remaining lifetimes, something that has not received much interest in the economics literature. I build on the fact that at any given chronological age the coefficient of variation of longevity (CoVoL) for low-income earners is larger relative to high-income earners. In some sense, life (financially) is both relatively and absolutely riskier for the poor. This then implies that their {\em willingness-to-pay} (WtP) for longevity insurance and the {\em annuity equivalent wealth} is greater, relative to high-income earners. 

Ergo, in a mandatory DB pension system there are two competing or opposite effects. On the one hand there is a clear and expected transfer of wealth from poor (i.e. higher mortality) to rich (i.e. lower mortality). On the other hand, economically disadvantaged participants benefit more from risk pooling due to the higher risk. This paper -- and in particular the main equation for $\delta$ -- helps locate the cut-off point.

If I can summarize with the characters I introduced at the very beginning, both Heather and Simon benefit from longevity risk pooling, that is owning life annuities, regardless of whether they are pooled (and swim) with people like themselves or forced to pool with others. When annuities are fairly priced, that is tailored to their own risk-type and we swim in small segregated pools, Simon derives relatively higher benefit from pooling longevity risk and holding annuities. Mother nature endows him with a higher mortality rate together with a slower mortality {\em growth} rate. This then is synonymous with (or leads to) a higher volatility of longevity and risk averse retirees are willing to pay (dearly) to mitigate the higher longevity risk. 

In some sense, nature's {\em Compensation Law of Mortality} leads to a higher demand for longevity insurance from those with the highest mortality rate. In contrast, Heather's relative benefit from annuitization might be (much) smaller than Simon's because her life expectancy is (much) higher, which means her mortality rate is lower and her individual volatility of longevity is lower. Practically speaking (and in the real world), after adding a relatively small insurance profit loading -- and perhaps some pre-existing annuity income to her portfolio -- Heather might derive no value from additional annuitization. Again, this is when she swims alone. Once they are forced into the same pool, although Simon is paying a penalty (implicit insurance loading) and is subsiding Heather, she now is benefiting from cheaper annuities. This increases her willingness to pay. So, whether the value of pooling is positive is an empirical question and the data presented indicate that (for now) it is still the case.

\clearpage

\begin{singlespace}

\end{singlespace}

\renewcommand\thesection{\Alph{section}}
\renewcommand\thesubsection{\thesection.\arabic{subsection}}
\setcounter{section}{0}
\section{Technical Appendix}

The purposes of this technical appendix is first: (A1.) carefully describe how to calibrate the Benjamin Gompertz (1825) law of mortality to any set of discrete mortality rates (or tables) via a linear least-squares methodology and then use a second least-squares to calibrate the ``compensation'' relationship; and then (A2) formally derive the annuity factor {\tt a}$(x)$ under a Gompertz law of mortality; then (A3.) sketch the derivation for the analytic expression for the {\em annuity equivalent wealth} (AEW) as a function of the above noted Gompertz parameters and wrap-up in (A4.) 

\subsection{Calibrating the Gompertz and CLaM Model}

The Compensation Law of Mortality (CLaM) postulates that for a heterogeneous group within a given species at a fixed chronological age, relatively healthy members with lower death rates (for example those with higher income) age faster, while those with higher death rates, who are sicker than average, age more slowly. The extreme form of the CLaM suggests that instantaneous mortality hazard rates converge to a constant at some advantaged age. 

To properly model this effect, I begin with a homogenous sub-group of the population under which each member is identified by two parameters: $h[i],g[i]$, where $i=1..N$, is the number of sub-groups. In our context, $i$ represents income percentiles ($N=100$) as in Chetty et al. (2016). Note that $h[i]$ represents a hypothetical age-zero biological hazard rate and $g[i]$ is the corresponding mortality growth rate, assuming no restrictions at this early stage other than: $h[i]>0$ and $g[i] \geq 0$. Also, despite the phrase age zero, I'm not modeling the early years of life (and I'm ignoring infant mortality.)

So, practically speaking, at any chronological age: $x > > 0$, the total mortality hazard rate: $\lambda_{x}[i]$, of members in sub-group $i$, obeys the so-called Gompertz-Makeham (GM) relationship up to some advanced age, after which it flattens out. Formally:
\begin{equation}
\lambda_{x}[i] \; = \; \left\{ \begin{array}{cl}
\lambda + h[i] e^{g[i]x} & x<x^{*}[i] \\
\lambda^{*}[i] & x \geq x^{*}[i]
\end{array} \right.
\label{EQ1}
\end{equation}

Note that the non-biological (and non-time dependent) hazard rate: $\lambda \geq 0$, is constant for for all members of the population, but $\lambda^{*}[i] >> \lambda$ and corresponding $x^{*},$ is group specific. 

Equation (\ref{EQ1}) is quite general. First, the plateau could depend on $i$, that is $\lambda^{*}[i] \neq \lambda^{*}[j]$, for $i \neq j$. Furthermore, for some $i$, it's conceivable $x^{*}[i] \rightarrow \infty$, and there is no (finite) mortality plateau. Rearranging equation (\ref{EQ1}), the GM model can also be expressed as:
\begin{equation}
\overbrace{\, \ln (\lambda_{x}[i] \; - \lambda) \, }^{Q_x} \; = \; \overbrace{\, \ln h[i] \, }^{C_0} + \overbrace{g[i]}^{C_1} x, \;\;\; \forall \, x<x^{*}[i],
\label{EQ2}
\end{equation}
which is the standard linear representation of (log) biological mortality rates for all ages in the ``Gompertzian" regime. Note that I deliberately use: $Q_x$, and not the standard one-year death rate: $q_x$, on the left-hand side, since they are not quite the same thing. More on this later. Chetty et al. (2016) assume that $Q_x=q_x$ and report estimated values for $C_0$ (the intercept) and $C_1$ (the slope), assuming that $\lambda=0$, based on U.S. mortality rates during the period 2001 to 2014, for ages $x$ up to 63. Milligan and Schirle (2018) do the same for Canadian data and have also obtained estimates for $C_0$ and $C_1$, with a similar structure. The values of $C_0$ and $C_1$ are then used to project or forecast one-year death rates at higher middle and higher ages (for which data is not available in their sample.) Generally speaking, they all find that the best fitting values of: $C_0$ have a negative relationship with the values of $C_1$. Figure \#5 and Figure \#6 illustrated this.

To get back to the distinction between $Q_x$ and $q_x$, in the GM framework the one-year death rate $q_x$, at any given chronological age $x$, is related to the continuous mortality rate, via:
\begin{equation}
1- q_x \; = \; e^{-\int_{x}^{x+1} \lambda_{y}dy}.
\end{equation}
When $\lambda_{x}=\lambda$, is constant (i.e. $h=0$), the survival rate to any time $t$ is $s(t)=e^{-\lambda t}$, and then $q_x = 1-e^{-\lambda},$ for any one year. In this (simplistic, clearly non-Gompertz) case, the parameter $\lambda_x$ is synonymous with a {\em continuously} compounded mortality rate and $q_x$ is the {\em effective} annual (one year) death rate. In the full Gompertz-Makeham ($h>0$) case, we have the following relationship between $q_x$, and the model parameters $(\lambda,h,g)$ for any sub-group:
\begin{eqnarray}
-\ln[1- q_x] 
= & \int_{0}^{1} \, (\, \lambda + h e^{g(x+s)} \, )ds \nonumber \\
= & \int_{0}^{1} \lambda ds + h e^{gx} \int_{0}^{1} e^{gs} ds  \nonumber \\
= & \lambda + h e^{gx} \left(e^g-1\right)/g
\end{eqnarray}

By definition: $-\ln[1- q_x] > \lambda \geq 0$, so we can subtract $\lambda$ from both sides, take logs (again) and obtain a linear relationship between the one-year death rate (on the LHS) and age $x$ (on the RHS), via the GM parameters:
\begin{equation}
\ln \left ( \ln \left( \frac{1}{1-q_x} \right) - \lambda \right) \; = 
\; \overbrace{\ln[h] \, + \, \ln [(e^g-1)/g]}^{K} 
+ g \, x,
\label{yx}
\end{equation}
where the new constant: $K$ is defined for convenience and suggests the proper regression (or least squares) methodology for calibrating: $\lambda,h,g$. The objective is to estimate $(\lambda, h,g)$ values from a vector of (empirical) $\tilde{q}_x$ values. To that end, we define a new variable:
\begin{equation}
\tilde{z}\; := \;  \ln \left ( \ln \left( \frac{1}{1-\tilde{q}_x} \right) - \lambda \right), 
\end{equation}
with the understanding that $\lambda \geq 0,$ is {\em known} and {\em fixed} in advance. For now, this leads to the (basic) Gompertz regression equation:
\begin{equation}
\tilde{z}_j\; = \; K \, + \, g x_j + \, \epsilon_j,
\label{eq5}
\end{equation}
where $x_j$ is a vector of ages, for example $x_1=35, x_2=36, x_3=37$, etc., and the $\tilde{z}_i$ are computed from the one-year death rates $\tilde{q}_{x_i}$. Running the regression formulated in equation (\ref{eq5}) leads to the best-fitting intercept and slope parameters $\tilde{K}$ and $\tilde{g}$, which based on equation (\ref{yx}), trivially result in unbiased estimates for the Gompertz parameters:
\begin{eqnarray}
g & = & \; \tilde{g}  \nonumber \\
\ln[h] & = & \tilde{K} - \ln[(e^{\tilde{g}}-1)/ \tilde{g}] \nonumber \\
h & = & e^{\tilde{K}} \left( \frac{\tilde{g}}{e^{\tilde{g}}-1} \right).
\label{eq6}
\end{eqnarray}
These, respectively, are the mortality growth rate, the log (biological) hazard rate and the actual (biological) hazard rate at age zero. In fact, knowing the Gompertz model leads to (extremely) high and significant coefficients, one is tempted to skip the formal regression (tests) and estimate $(h,g)$ via the equation for the least-squares line:
\begin{equation}
C_1 = \frac{\sum_{1}^{N}(x_j-\bar{x})(y_j-\bar{y})}{\sum_{1}^{N}(x_j-x)^2} \; , \; C_0 = \bar{y}-C_1\bar{x},
\end{equation}
where $\bar{y}$ and $\bar{x}$ are the arithmetic mean of $y$, and $x$ respectively. And, since the age variable is a linear sequence: $\bar{x} = (x_{{\tt min}}+x_{{\tt max}})/2$. In sum, regardless of the exact calibration methodology, the above procedure leads to a pair of values: $(h,g)$ for every sub-group $i$, in the population.

Now, the weak-form {\em Compensation Law of Mortality} states that groups with relatively higher biological hazard rates: $h[i] > h[j]$, experience relatively lower growth rates $g[i] < g[j]$, and vice versa. See Gavrilov \& Gavrilova (original article 1979, book 1991) for more. In other words, the CLaM posits a formal analytic relationship between $h[i]$ and $g[i]$, denoted by $h:=h(g)$, within a range of: $g_{\tt min} \leq g \leq g_{\tt max}$. To be clear, the weak form CLaM (only) stipulates that: $\partial h(g) / \partial g < 0$, if one thinks of $h$ as a function of $g$.

A strong-form CLaM begins at the very end of the lifecycle by postulating that: $x^{*}[i]=x^{*}, \forall i$, and the mortality plateau is identical for all sub-groups. This actually places much tighter restrictions on the function $h(g)$, and by equation (\ref{EQ2}) implies: 
\begin{equation}
L \; := \; \ln (\lambda^{*} \; - \lambda) \; = \; \ln h(g) + g x^{*},
\label{EQ3}
\end{equation}
where $L$ is a (new) convenient constant. Rearranging equation (\ref{EQ3}) leads to a linear representation for the function: $\ln h(g)$, and can be expressed as:
\begin{equation}
\ln h(g) \; = \; L \; - x^{*} \, g,
\label{EQ4}
\end{equation}
I will refer to and label: $\ln h(g)$, as the CLaM function. Exponentiating equation (\ref{EQ3}), the actual age-zero biological hazard rate: $h(g)$ can be expressed as:
\begin{equation}
h(g) \; = (\lambda^{*} - \lambda) e^{-x^{*}g},
\label{EQ5}
\end{equation}
which (at $g=0$) recovers the mortality plateau: $\lambda^{*} = h(0)+\lambda$. So, under the strong compensation law of mortality, I can rewrite equation (\ref{EQ1}) as:
\begin{equation}
\lambda_{x}(g) \; = \; \left\{ \begin{array}{cl}
\lambda + (\lambda^{*}-\lambda) e^{g(x-x^{*})} & x<x^{*} \\
\lambda^{*} & x \geq x^{*}
\end{array} \right.
\label{EQ6}
\end{equation}
On to estimation. Recall that I have access to a set of 100 values of: $\{\ln h[i],g[i]\}$, and {\em assuming} they are consistent with the strong form of the CLM, I can estimate the (intercept) $L$, and (slope) $x^{*}$ via regression. In particular, as per equation (\ref{EQ4}), the relationship is:
\begin{equation}
\overbrace{\; \ln h[i] \; }^{y_j} \; = \; \overbrace{L}^{C_0} \; + \; \overbrace{(-x^{*})}^{C_1} \, \overbrace{g[i]}^{z_j} \, + \epsilon_j.
\label{EQ7}
\end{equation}
Note that this 2nd regression procedure shouldn't be confused with the first regression procedure that is used to extract or estimate the original Gompertz parameters in equation (\ref{EQ2}). The first regression (or least squares estimation) is what leads to the: $\ln h[i]$ and $g[i]$ values, as in Chetty et al. (2016) or Milligan and Schirle (2018). 

To be clear, once we actually have the $(h,g)$ parameters, which will be used to price annuities and value pooling, the point of the second procedure is to confirm the existence of a CLaM-effect, which implies higher CoVaL for lower incomes and vice versa.

\begin{center}
\fbox{TABLE \#5 GOES HERE}
\end{center}

In sum, table \#5 displays the results of what I call the 2nd regression which (in a sense) tests for the presence of a strong CLaM in the data. Indeed, the relationship between: $\ln h[i]$ and $g[i]$ is linear with $R^2$ values close to 98\%, providing support for a strong version CLaM within the U.S. population segmented by income. The estimated $L = \ln(\lambda^{*}-\lambda)$, reveals or locates the plateau. And, the slope $(-x^{*})$ is the age at which it's achieved, a.k.a. the {\em species specific lifespan}, per Gavrilov and Gavrilova (1979). 

As a final side note, there is theoretical support for: $L \approx \ln (\ln 2))$, so that one year survival and mortality rates at the plateau are: $e^{-\lambda^{*}}=0.5=e^{-e^L}$, assuming the accidental mortality rate (a.k.a. the Makeham constant) is zero. The Chetty et al. (2016) indicate lower values for $L$, and so do the Milligan and Schirle (2018) data.

\subsection{Closed-form Annuity Factors and Moments}

Note that within each population sub-group, and up to the mortality plateau, once can express the continuous-time {\em hazard rate} function as:
\begin{equation}
\lambda_{x+t} \; = \; h_{x} \, e^{gt} \; = \; \frac{1}{b} e^{(x+t-m)/b},
\end{equation}
where $h_{x}$ is the hazard rate at (an arbitrary baseline) age $x$, and $g$ is the hazard {\em growth} rate. Recall that the parameter $g$ represents the slope in a regression of {\em log} hazard rates (as an independent variable) on age (the dependent variable), as done in the prior sub-section. Note again that: $b=1/g,$ which measures the {\em dispersion} of the remaining lifetime random variable denoted by: $T_x$ in the $(m,b)$ formulation. 

For what follows here I work with the: $(h_x,g)$ formulation, where $h_x$ is the mortality hazard rate at the current age $x$, which leads to a cleaner and more intuitive relationship with the annuity equivalent wealth $\delta$. One can easily move from $(m,b)$-space to $(h_x,g)$-space. For example, fixing the hazard rate at age 65: $h_{65}=0.5\%$ and assuming a growth rate of: $g=10\%$, the modal value of the lifetime is: $m=94.957=65-\ln[0.005/0.1]/0.1$ years and the dispersion value is: $b=1/g=10$. Likewise, if: $h_{65}=0.5\%$, but the growth rate is $g=8\%$, the: $m=98.761$ years and $b=12.5$ years.

Moving on, under the $(h_x,g)$ hazard rate formulation, the conditional survival probability, denoted by: $p(t,h_x,g)$ is equal to:
\begin{equation}
p(t,h_x,g) \; = \; \exp \{ -\int_0^{t} \, h_{x+s} \, ds \} \, = \, \exp \{ (h_x/g) (1-e^{gt}) \}.
\end{equation}
Recall that any immediate life annuity factor can be expressed as:
\begin{eqnarray}
{\tt a}(r,h_x,g) \; = & \; \int_{0}^{\infty} \; e^{-rt} \, p(t,h_x,g) \, dt \\
= & \; \int_{0}^{\infty} \; e^{-rt} \exp \left \{ (h_x/g) (1-e^{gt}) \right \} dt \;
\nonumber \\
= & \; e^{h_x/g} \int_{0}^{\infty} \; e^{-rt} \exp \left \{ - (h_x/g) e^{gt} \right \} dt \; \nonumber 
\end{eqnarray}
I can simplify the integral using a change of variables: $s=(h_x/g) e^{gt}$, therefore:
\begin{equation}
s=\frac{h_x}{g} e^{gt}  \hspace{0.2in} \rightarrow \hspace{0.2in} t = \ln{[\frac{sg}{h_x}]}/g 
\hspace{0.20in} \rightarrow \hspace{0.20in}  dt= \frac{1}{sg} ds
\label{change.variable}
\end{equation}
Using the new variable $s$, instead of $(h_x/g) e^{gt}$, we can simplify the integrand: 
\begin{equation}
 e^{h_x/g} \int_{0}^{\infty} \; e^{-rt} \exp \left \{ - (h_x/g) e^{gt} \right \} dt =  \; e^{h_x/g} \int_{0}^{\infty} \; e^{-rt} e^{-s} dt 
\end{equation}
We can now replace the $t$ with: $ \ln{[sg/h_x]}/g$ to obtain:
\begin{equation}
e^{h_x/g} \int_{0}^{\infty} \; e^{-rt} e^{-s} dt =  \; e^{h_x/g}\int_{0}^{\infty} \; e^{-r(\ln{[\frac{sg}{h_x}]}/g )} e^{-s} dt \; = \; e^{h_x/g} \int_{0}^{\infty} \; e^{-s} (\frac{sg}{h_x})^{(\frac{-r}{g} )} dt 
\end{equation}
Replacing $dt$ with $(1/sg)ds$ and moving all non $s$ terms outside the integral:
\begin{equation}
{\tt a}(r,h_x,g) = \; e^{h_x/g} \int_{0}^{\infty} \; e^{-s} (\frac{sg}{h_x})^{(\frac{-r}{g} )} \frac{1}{sg} ds \;
=  \; e^{h_x/g} (\frac{g}{h_x})^{(\frac{-r}{g} )} \frac{1}{g}\int_{h_x/g}^{\infty} \; e^{-s} s^{(\frac{-r}{g} )-1} ds
\end{equation}
The terms outside and to the left the integral can be simplified to:
\begin{equation}
e^{h_x/g} (\frac{g}{h_x})^{(\frac{-r}{g} )} \frac{1}{g} \; = \;  \frac{1}{g} \, \exp\{ \frac{1}{g} \, (h_x+r \, \ln{[h_x/g]})\}
\; = \; \frac{1}{g \, \exp\{(-1/g) \, (h_x + r \ln{[h_x/g]} \,  \}}
\end{equation}
The Gompertz (albeit without the Makeham constant) life annuity factor can now be re-written formally as:
\begin{equation}
{\tt a}(r,h_x,g)  \;=  \; \frac{1}{g \, \exp\{(-1/g) \, (h_x + r \ln{[h_x/g]} \,  \}} \int_{h_x/g}^{\infty} \; e^{-s} s^{(\frac{-r}{g} )-1} ds
\label{pre_gamma}
\end{equation}
From the messy looking equation (\ref{pre_gamma}) it might not {\em appear} as if I have improved matters, but the integral can actually be identified as the Incomplete Gamma (IG) function:
\begin{equation}
\Gamma(\alpha,\beta) \, = \, \int^{\infty}_{\beta} \, e^{-s} \, s^{\alpha-1} ds.
\label{IncG}
\end{equation}
When the lower bound of integration $\beta=0$ the IG function collapses to the basic Gamma function and when $\alpha$ is an integer  then $\Gamma(\alpha,0)=(\alpha-1)(\alpha-2)...$ etc., a.k.a. $(\alpha-1)$ factorial, with the understanding that both $\Gamma(1,0)=1$ and $\Gamma(2,0)=1$. 

For general values of $\alpha$ and $\beta$ the IG function is readily available in most business and scientific software packages (as well as R, of course), similar to the {\em error function} or the normal distribution. For example, the value of $\Gamma(-0.5,1)=0.178148$ to five digits and the value of: $\Gamma(-0.5,0.3678)=0.89635$. I do caution that for non-positive values of $\alpha$ there are some numerical stability issues.
Merging equation (\ref{IncG}) and (\ref{pre_gamma}), I can write the annuity factor using a closed-form expression:
\begin{equation}
{\tt a}(r,h_x,g) \; = \frac{\Gamma(-r/g,h_x/g)}{g \, \exp\{(-1/g) \, (h_x + r \ln{[h_x/g]})\}},
\label{GAVM1}
\end{equation}
This is our basic annuity factor. Here are some numerical examples. Let's arbitrarily set the interest rate $r = 3\%$. If we keep $g$ constant at $0.08$, here is what happens when $h_x$ takes values $0.1$ , $0.2$, and $0.3$, respectively.\ ${\tt a}(0.03,0.1,0.08) = 5.552432$,  ${\tt a}(0.03,0.2,0.08) = 3.464195$, and: ${\tt a}(0.03,0.3,0.08) =  2.543422$ Note that as $h_x$ increases, the value of the annuity factor declines. This should be intuitive because the force of mortality (or hazard rate) kills you, so if it increases you will live a shorter life and thus receive less income, making the annuity factor cheaper. Likewise, if we fix: $h_x=0.1$ and change $g$ to take values $0.09$, $0.12$, $0.15$, respectively. Then, ${\tt a}(0.03,0.1,0.09) = 5.392625$, and: ${\tt a}(0.03,0.1,0.12) = 4.981276$, and finally: ${\tt a}(0.03,0.1,0.15) = 4.646376$, all in dollars. So, as $g$ increases, the value of the annuity factor declines, while holding $h_x$ constant. 

\begin{center}
\fbox{FIGURE \#7 GOES HERE}
\end{center}

Figure \#7 shows this graphically. The x-axis represents the current (or initial) hazard rate $h$, ranging from $h_x=0.005$ to $h_x=0.5$ in equal increments.  This happens to correspond to a range of: $x=64$ to $x=101$, assuming $m=90$ and $b=8$, in the alternate $(m,b)$ specification. The x-axis is labeled with both a rate (top) and an age (below). And, while the former (top row) increases linearly, the latter (bottom row) does not, since $x:= b\ln[hb]+m$ under the Gompertz law. The y-axis in figure \#7a represents the annuity factor corresponding to that particular (x-axis) hazard rate and age, assuming: $g=12.5\%$ in the above $(h,g)$ formula. Notice how the annuity factor declines, as we move from left to right and the hazard rate (as well as the age) increases. Panel \#7b plots the value of the annuity factor over the same range of $h_x=0.005$ to $h_x=0.5$, assuming a reduced: $g=8\%$. 

The annuity factor is higher at every value (as one can see from the few points that are highlighted), although it also declines in hazard rate and/or age. Note that in this case the corresponding $b=1/g=12.5$ years, and the corresponding (second) x-axis values range from $x=55$ to $113$. Finally, recall that under the Gompertz $(m,b)$ formulation, common in the actuarial finance, the mortality hazard rate is expressed as: $\lambda_{x+t}=(1/b)e^{(x+t-m)/b}$. In that case, the immediate annuity factor {\tt a} can be derived in terms of $(x,m,b)$ by substituting: $h_x$ with $(1/b)\exp\{\frac{x-m}{b}\}$ and replacing $g$ with $1/b$. For completeness I present:
\begin{equation}
{\tt a}(r, x,m,b)  \; = \frac{\; b \Gamma(-rb,\exp\{\frac{x-m}{b}\})}{\exp\{r(m-x) \, - \, \exp\{ \frac{x-m}{b} \} \}}.
\label{GAVM2}
\end{equation}

\subsection{Deriving the Annuity Equivalent Wealth or $\delta$}
I sketch a quick proof of the expression for the annuity equivalent wealth (AEW): $1+\delta$, under the assumption the individual has {\bf no} pre-existing annuity income. This is based on the derivation in Milevsky and Huang (2018), who provide various closed-form expressions for the AEW under alternate mortality assumptions and more general (non-zero) pension income. See also Cannon and Tonks (2008). Let $u(c)$ denote a constant relative risk aversion (CRRA) utility (a.k.a. felicity) function parameterized by risk aversion $\gamma$, and a subjective discount rate $\rho=r$. Formally, $u(c)=c^{1-\gamma}/(1-\gamma)$. The maximal utility {\em without} annuities is: 
\begin{equation}
U^{*}(w) \; = \max_{c_t} \; \int_{0}^{\omega-x} \, e^{-r t} \, p(t,h^i,g^i) \, u(c_t) \, dt,
\label{eq2}
\end{equation}
where $(h^i,g^i)$ are the Gompertz parameters at the relevant income percentiles, and the budget constraint is:
\begin{equation}
dW_{t} \; = \; (rW_{t} \, - \, c_t) \, dt, \; \; W_{0} = w.
\label{eq3}
\end{equation}
The optimal consumption function is denoted by: $c^{*}_{t}$, and do not allow any borrowing so that wealth $W_t \geq 0$ at all times. The only reason to prefer early vs. late consumption, is due to mortality beliefs and the inter-temporal elasticity of substitution, $1/\gamma$. Now, since there is no pre-existing pension income, the relevant consumption rate must be sufficient to last all the way to $\omega$, so that:
\begin{equation}
w=c^{*}_{0} \int_0^{\omega-x} e^{-rt}p(t,h^i,g^i)^{1/ \gamma}dt,
\end{equation}  
which leads to the corresponding:
\begin{equation}
c^{*}_t \; = \; \left( \frac{w}{ \int_{0}^{\infty} e^{-rt}p(t,h^i,g^i)^{1/ \gamma} dt  } \right) p(t,h^i,g^i)^{1 / \gamma},
\label{c}
\end{equation} 
The integral in the denominator of equation (\ref{c}) is an annuity factor (of sorts) assuming the survival probability $p(t,h^i,g^i)$ is shifted by $1/ \gamma$. For example, when $\gamma=1$, the optimal consumption function $c^{*}_{t}$ in equation (\ref{c}) collapses to the hypothetical annuity: $w/{\tt a}(r,h^i,g^i)$ times the survival probability $p(t,h^i,g^i)$, which is less than what a life annuity would have provided. The individual who converts all liquid wealth $w$ into the annuity would consume $w/${\tt a}, but the non-annuitizer reduces consumption in proportion to survival probabilities. In contrast to: $U^{*}(w)$, let $U^{**}(w)$ denote discounted lifetime utility of wealth, assuming wealth $w$ {\bf is entirely} annuitized or pooled at age $x$. Discounted utility is:
\begin{equation}
U^{**}(w)\; = \; \int_{0}^{\infty} \, e^{-rt} \, p(t,h^i,g^i) \, u(w/{\tt a}(r,\hat{h},\hat{g})) \, dt,
\label{u3}
\end{equation}
where the optimized consumption path is trivially $c^{*}_t=w/{\tt a}(r,\hat{h},\hat{g})$, for all $t$. The next (and essentially final) step is to note that $\delta$ will satisfy the following equation:
\begin{equation}
U^{*}((1+ \delta_i)w)  \; = U^{**}(w),
\end{equation}
as per the definition of AEW. We refer the interested reader to Milevsky and Huang (2018) for the algebra that extracts: $\delta_i$ from the above equation.

\subsection{Comparative statics for $\delta$}

The final figure (\#8) brings the entire technical appendix together in one summary picture. It plots the equation for the value of longevity risk pooling: $\delta_x$, for a range of mortality growth rates $g$, assuming four different hypothetical mortality hazard rates $h_x$. The top row (red) dots fixes the initial mortality rate at 3\%, the second row (blue) assumes it's 2\%, the third row (purple) is based on 1\% and the bottom row (black) is for 0.5\%. These four mortality hazard rates are comparable with (although not exactly equal to) the range of numbers for high-income versus low-income displayed in table \#4, assuming $r=3\%$, $\gamma=3$.

\begin{center}
\fbox{FIGURE \#8 GOES HERE}
\end{center}

Notice that at any specific value of mortality growth rate $g$ on the x-axis, the so-called high mortality $\delta$ values sit uniformly on top of the low mortality $\delta_x$ values. But, as one moves from left to right and increases the value of $g$, all else being equal, the impact on $\delta_x$ is non monotonic. Indeed, $\partial \delta_x / \partial g$ can't be signed. At low mortality rates it actually increases (in $g$) and at high mortality it declines (in $g$). However, {\bf if} I superimpose the suitably-calibrated {\em compensation law of mortality} line on this plot, and only focus or use biologically realistic combinations of $h_x$ and $g$, i.e. where the arrow pierces the dots, a clear pattern emerges. Increasing the mortality growth rate $g$ forces down the initial mortality rate $h_x$ itself (per CLaM). The value of longevity risk pooling declines in $g$ and therefore increases in $1/g$. Hence, I conclude that the value of $\delta_x$ increases in both the coefficient of variability of longevity (CoVoL) as well as the standard deviation of longevity, thanks to CLaM. {\bf Q.E.D.}


\def\baselinestretch{1.0}

\clearpage
\begin{table}
\begin{center}
\begin{tabular}{||c|c|c||} \hline \hline
\multicolumn{3}{||c||}{{\bf Table \#1: Intuition for Pension Subsidies}} \\ \hline\hline
 & {\bf {\color{red} Simon}} & {\bf Heather}  \\ \hline
{\bf Current Age:} & 65 & 65 \\ \hline
{\bf Earned Pension Credits:} & {\em Maximum} & {\em Maximum} \\ \hline
{\bf Annual Pension Income:} & \$25,000 & \$25,000 \\ \hline
{\bf Life Expectancy (Horizon):} & {\color{red} 10 years} & 30 years \\ \hline 
{\bf Term Annuity PV at 3\%:} & \$212,750 & \$487,250 \\ \hline
{\bf Funded System Assets:} & \multicolumn{2}{|c||}{\$700,000} \\ \hline
{\bf Total Pension Contributions:} & \$350,000 & \$350,000 \\ \hline
{\bf Transfer \& Subsidy:} & {\color{red} -\$137,250} & +\$137,250 \\ \hline \hline
\end{tabular}
\end{center}
\end{table}
Note. See body of paper and in particular the introduction for details and context.

\clearpage
\begin{table}
\begin{center}
\begin{tabular}{||c|c|c|c|c|c|c||}
\hline\hline
\multicolumn{7}{||c||}{{\bf Table \#2a: U.S. Death Rates per 1,000 individuals.}} \\ \hline\hline
  & \multicolumn{3}{|c|}{{\bf MALE}} & \multicolumn{3}{|c||}{{\bf FEMALE}} \\ \hline\hline
{\bf Income Group} & {\bf Age=40} & {\bf Age=50} & {\bf Age=60} & {\bf Age=40} & {\bf Age=50} & {\bf Age=60} \\ \hline\hline  
Lowest (1st pct.) & 5.8 & 12.5 & 22.1 & 4.3 & 8.0 & 12.8 \\ \hline
25th percentile & 2.0 & 4.5 & 10.9 & 1.2 & 2.7 & 5.9 \\ \hline
Median (50th pct.) & 1.2 & 2.9 & 7.3 & 0.8 & 2.0 & 4.5 \\ \hline
75th percentile & 0.8 & 1.8 & 4.9 & 0.5 & 1.3 & 3.5 \\ \hline
Highest (100th pct.) & 0.6 & 1.1 & 2.8 & 0.3 & 0.8 & 2.2 \\ \hline \hline
\end{tabular}
\label{tab1}
\smallskip
\end{center}
\end{table}

\begin{table}
\begin{center}
\begin{tabular}{||c|c|c|c|c||}
\hline\hline
\multicolumn{5}{||c||}{{\bf Table \#2b: Mortality Growth Rate and Projections}} \\ \hline 
{\bf Income} & \multicolumn{2}{|c|}{{\bf Male}} & \multicolumn{2}{|c||}{{\bf Female}} \\ \hline
{\bf Group} & $g:=\frac{\ln q_{x+T}}{\ln q_{x}}/T $ & $1000\tilde{q}_{100}$ & $g:=\frac{\ln q_{x+T}}{\ln q_{x}}/T $  & $1000\tilde{q}_{100}$  \\ \hline\hline
Lowest (1st pct.)  & 5.63\% & 208.5 & 4.81\% & 89.2 \\ \hline
25th Percentile  & 8.74\% & 355.2 & 8.32\% & 171.1 \\ \hline
Median (50th pct.) & 9.21\% & 285.7 & 8.68\% & 158.9 \\ \hline
75th Percentile  & 10.22\% & 302.1 & 10.21\% & 211.4 \\ \hline
Highest (100th pct.) & 10.00\% & 163.2 & 9.97\% & 115.0  \\  \hline
\end{tabular}
\label{tab2}
\smallskip
\end{center}
\end{table}
Note: The source is Chetty et al. (2016) with Mortality over the period: 2001 to 2014 using an income lag of two years. Calculation (by author) of $g$ in the second panel was based on growth from age $x=50$ to $x=63$. Each percentile's $g$ value was used to forecast: $\tilde{q}_{100}$. Note that by age $x=100$, projected mortality rates are within a multiple of two of each other. According to theory they should converge.

\clearpage
\begin{table}
\begin{center}
\begin{tabular}{||c|c|c|c|c|c|c|c|c||}
\hline\hline
\multicolumn{9}{||c||}{{\bf Table \#3: Coefficient of Variation of Longevity (CoVoL)}: $\varphi_x$} \\ \hline
\multicolumn{9}{||l||}{...when modal value of life $(m)$ is: {\bf 98 years}.} \\ \hline\hline
{\bf Mortality} & {\bf Disp.} & {\bf Mean} & {\bf StDev.} & {\bf CoVoL} & {\bf Hazard} & {\bf Mean} & {\bf StDev.} & {\bf CoVoL} \\ \hline\hline
{\bf Growth} & $b$ & $E[T_0]$ & $SD[T_0]$  & $\varphi_0$ & $\lambda_{65}$ & $E[T_{65}]$ & $SD[T_{65}]$ & $\varphi_{65}$ \\ \hline\hline
11.5\% ({\bf rich}) & 8.696	& 92.98	& 11.15	& 12.0\%	& 0.2586\% & 28.82	& 9.70   & {\bf 33.7\%} \\ \hline
10.5\%	& 9.524	& 92.51	& 12.20	& 13.2\%	& 0.3284\% & 28.68	& 10.29	& 35.9\% \\ \hline
9.5\%	& 10.526	& 91.93	& 13.47	& 14.7\%	& 0.4132\%  & 28.59	& 10.95	& 38.3\% \\ \hline
8.5\%	& 11.765	& 91.23	& 15.00	& 16.4\%	& 0.5143\% & 28.59	& 11.69	& 40.9\% \\ \hline
7.5\%	& 13.333	& 90.37	& 16.90	& 18.7\%	& 0.6312\% & 28.72	& 12.55	& 43.7\% \\ \hline
6.5\%	& 15.385	& 89.30	& 19.26	& 21.6\%	& 0.7609\%  & 29.08	& 13.58	& 46.7\% \\ \hline
5.5\%	& 18.182	& 87.99	& 22.21	& 25.2\%	& 0.8956\% & 29.83	& 14.89	& 49.9\% \\ \hline
 \multicolumn{9}{||l||}{...when modal value of life $(m)$ is: {\bf 78 years}.} \\ \hline\hline
{\bf Mor.Gro} & $b$ & $E[T_0]$ & $SD[T_0]$ & $\varphi_0$ & $\lambda_{65}$ & $E[T_{65}]$ & $SD[T_{65}]$ & $\varphi_{65}$ \\ \hline\hline
11.5\%	& 8.696	& 72.99	& 11.11	& 15.2\%	& 2.5789\% & 12.30	& 6.55	& 53.3\% \\ \hline
10.5\%	& 9.524	& 72.53	& 12.14	& 16.7\%	& 2.6815\% & 12.64	& 6.90	& 54.6\% \\ \hline
9.5\%	& 10.526	& 71.97	& 13.35	& 18.5\%	& 2.7629\% & 13.08	& 7.33	& 56.0\% \\ \hline
8.5\%	& 11.765	& 71.32	& 14.79	& 20.7\%	& 2.8153\% & 13.66	& 7.85	& 57.5\% \\ \hline
7.5\%	& 13.333	& 70.55	& 16.51	& 23.4\%	& 2.8289\% & 14.43	& 8.51	& 59.0\% \\ \hline
6.5\%	& 15.385	& 69.65	& 18.57	& 26.7\%	& 2.7921\% & 15.49	& 9.36	& 60.4\% \\ \hline
5.5\% ({\bf poor}) & 18.182 & 68.69	& 21.07	& 30.7\%	& 2.6906\% & 17.00	& 10.53   & {\bf 61.9\%} \\ \hline\hline
\end{tabular}
\label{tab3}
\smallskip
\end{center}
\end{table}
Note: These numbers (and ranges) are for illustrative purposes and based on theoretical Gompertz values and do not correspond to any specific income percentile or group. Note that the standard deviation (SD) of longevity and the coefficient of variation of longevity (CoVoL) are both higher when the mortality growth rate $g$ increases. 

\clearpage
\begin{table}
\begin{center}
\begin{tabular}{||c|c|c|c|c|c|c||}
\hline\hline
\multicolumn{7}{||c||}{{\bf Table \#4: Annuity Equivalent Wealth (AEW) and Value of Risk Pooling:}} \\ \hline\hline
\multicolumn{5}{||c|}{{\bf Demographics}}  & \multicolumn{2}{|c||}{{\bf Risk Aversion:} $\gamma=3$}  \\ \hline 
{\bf Income} & {\bf Gompertz} & $E[T_{65}]$ & $\varphi_{65}$ & {\bf Annuity} & $\delta_{65}:=$ {\bf AEW-1}  & $\delta_{65}:=$ {\bf AEW-1}  \\ 
{\bf Percentile} & $(h_{65},g)$ & {\bf Years} & {\bf CoVoL} & {\bf Factor \$} & {\bf Individual}  &  {\bf \; Group \;} \\ \hline \hline
\multicolumn{7}{||c||}{{\bf FEMALE}} \\ \hline 
{\bf Lowest} &  $ (1.64\% , \, 5.29\%) $ & 22.75 & 56.73\% & 15.23 & 62.18\% & 46.52\% \\ \hline
5th &  $ (1.22\% , \, 6.68\%) $ & 23.36 & 51.05\% & 15.72 & 53.59\% & 43.18\% \\ \hline
10th &  $ (1.15\% , \, 8.08\%) $ & 21.52 & 48.63\% & 14.97 & 52.45\%  & 35.39\% \\ \hline
20th &  $ (1.00\% , \, 8.9\%) $ & 21.58 & 46.33\% & 15.08 & 49.20\% & 33.46\% \\ \hline
30th &  $ (0.86\% , \, 8.61\%) $ & 23.41 & 45.24\% & 15.96 & 45.87\% &  38.06\% \\ \hline
40th &  $ (0.78\% , \, 8.84\%) $ & 23.94 & 44.11\% & 16.23 & 43.93\% & 38.57\% \\ \hline
{\bf 50th} &  $ (0.69\% , \, 8.73\%) $ & 25.31 & 43.10\% & 16.86 & {\bf 41.46\%} & {\bf 41.46\%} \\ \hline
60th &  $ (0.69\% , \, 10.06\%) $ & 23.13 & 41.91\% & 15.94 & 41.89\% & 34.20\% \\ \hline
70th &  $ (0.55\% , \, 9.08\%) $ & 26.68 & 40.95\% & 17.51 & 37.82\% & 43.18\% \\ \hline
80th &  $ (0.50\% , \, 10.35\%) $ & 25.30 & 39.15\% & 17.00 & 36.81\% & 37.97\% \\ \hline
90th &  $ (0.45\% , \, 10.49\%) $ & 26.05 & 38.15\% & 17.36 & 35.12\% & 39.12\% \\ \hline
95th &  $ (0.38\% , \, 9.74\%) $ & 28.89 & 37.46\% & 18.53 & 32.44\% & 45.61\% \\ \hline
{\bf Highest} &  $ (0.34\% , \, 9.89\%) $ & 29.72 & 36.46\% & 18.89 & 30.89\% & 46.66\% \\ \hline
\multicolumn{7}{||c||}{{\bf MALE}} \\ \hline 
{\bf Lowest} &  $ (3.02\% , \, 6.56\%) $ & 14.76 & 61.24\% & 11.15 & 84.26\% & 38.25\% \\ \hline
5th &  $ (2.10\% , \, 6.63\%) $ & 17.93 & 56.97\% & 12.96 & 70.29\% & 48.52\% \\ \hline
10th &  $ (2.00\% , \, 7.46\%) $ & 17.37 & 55.13\% & 12.72 & 68.20\% & 43.96\% \\ \hline
20th &  $ (1.75\% , \, 8.31\%) $ & 17.55 & 52.56\% & 12.89 & 63.64\% & 41.90\% \\ \hline
30th &  $ (1.50\% , \, 8.78\%) $ & 18.29 & 50.41\% & 13.33 & 59.15\% & 42.76\% \\ \hline
40th &  $ (1.18\% , \, 8.43\%) $ & 20.82 & 48.41\% & 14.65 & 52.95\% & 50.76\% \\ \hline
{\bf 50th} &  $ (1.06\% , \, 8.83\%) $ & 21.16 & 46.97\% & 14.86 & {\bf 50.53\%} & {\bf 50.53\%} \\ \hline
60th &  $ (0.89\% , \, 8.68\%) $ & 23.01 & 45.45\% & 15.77 & 46.54\% & 55.47\% \\ \hline
70th &  $ (0.82\% , \, 9.31\%) $ & 22.75 & 44.08\% & 15.70 & 45.01\% & 53.18\% \\ \hline
80th &  $ (0.70\% , \, 9.49\%) $ & 23.81 & 42.58\% & 16.23 & 42.14\% & 55.20\% \\ \hline
90th &  $ (0.60\% , \, 9.8\%) $ & 24.68 & 40.98\% & 16.67 & 39.45\% & 56.39\% \\ \hline
95th &  $ (0.51\% , \, 9.68\%) $ & 26.26 & 39.83\% & 17.38 & 36.87\% & 60.09\% \\ \hline
{\bf Highest} &  $ (0.42\% , \, 8.74\%) $ & 30.13 & 39.01\% & 18.93 & 33.24\% & 69.77\% \\ \hline\hline
\multicolumn{7}{||c||}{Assumes $r=3\%.$ No explicit insurance loading, other than when paying group rates.} \\ \hline 
\end{tabular}
\label{tab4}
\smallskip
\end{center}
\end{table}

\clearpage
\begin{table}
\begin{center}
\begin{tabular}{||c||c|c|c|c|c|c||}
\hline\hline
\multicolumn{7}{||c||}{{\bf Table \#5: CLaM Regression with all Income Percentiles}} \\ \hline
{\bf Variable} & \multicolumn{3}{|c|}{{\bf MALE}} & \multicolumn{3}{|c||}{{\bf FEMALE}} \\ \hline
 & Coeff. & Std.Er & t-val. & Coeff. & Std.Er & t-val.  \\ \hline
Intercept $(L)$ & -1.234 & 0.119 & -10.4 & -2.038 & 0.129 & -15.8 \\ \hline
Slope: $(-x^{*})$ & {\bf -99.98} & 1.284 & -77.9 & {\bf -95.19} & 1.359 & -70.1 \\ \hline
Adj. $R^2$ & \multicolumn{3}{|c|}{98.39\%} & \multicolumn{3}{|c||}{98.02\%} \\ \hline
Plateau& \multicolumn{3}{|c|}{$\lambda^{*} \in [0.2585, \, 0.3278]$} & \multicolumn{3}{|c||}{$\lambda^{*} \in [0.1145, \, 0.1482]$} \\ \hline
Range: $g[i]$ & \multicolumn{3}{|c|}{$(5.961\%,\,10.491\%)$} & \multicolumn{3}{|c||}{$(5.454\%,\,10.621\%)$} \\ \hline
Mean: $g[i]$ & \multicolumn{3}{|c|}{$9.181\%$} & \multicolumn{3}{|c||}{$9.415\%$} \\ \hline
\end{tabular}
\end{center}
\end{table}
Note: As described in the appendix, these are the results from regressing the (male and female) Gompertz mortality intercepts on the mortality growth rates from the Chetty et al. (2016) dataset. Practically speaking, as the mortality growth rate $g$ increases, the implied age-zero mortality rate $h$ declines, reducing the value of homogenous pooling.


\clearpage
\begin{figure}
\vspace{-0.2in}
\caption{Remaining lifetime, dispersion and coefficient of variation of longevity (CoVoL).}
\begin{center}
\includegraphics[width=0.95\textwidth]{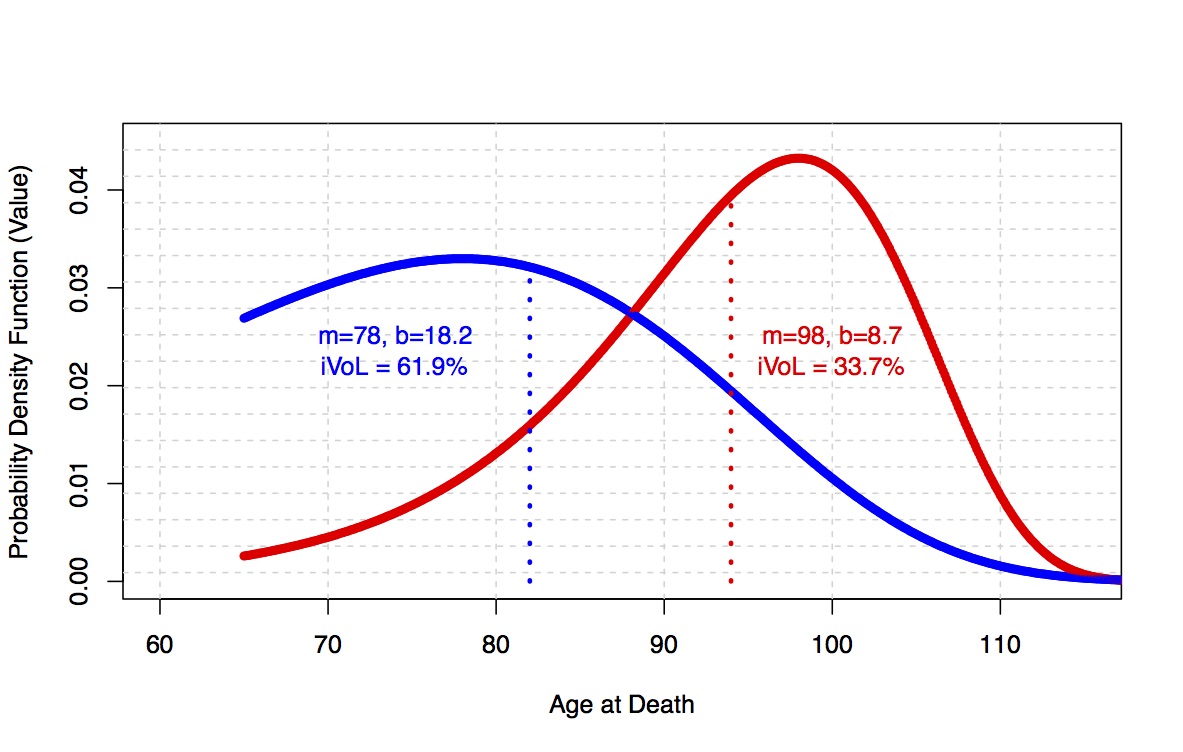} 
\end{center}
\end{figure}
Note: Both curves are based on the Gompertz law of mortality, conditional on age $x=65$. The right curve represents a retiree with a higher life expectancy ($m=98$) and lower dispersion parameter ($b=8.7$), leading to a coefficient of variation of longevity of 33.7\%. The left curve represents a retiree with lower life expectancy ($m=78$) and higher dispersion ($b=18.2$), whose CoVoL is double, at: 61.9\%.

\clearpage
\begin{figure}
\vspace{-0.2in}
\caption{Visualizing the Compensation Law of Mortality}
\begin{center}
\includegraphics[width=0.48\textwidth]{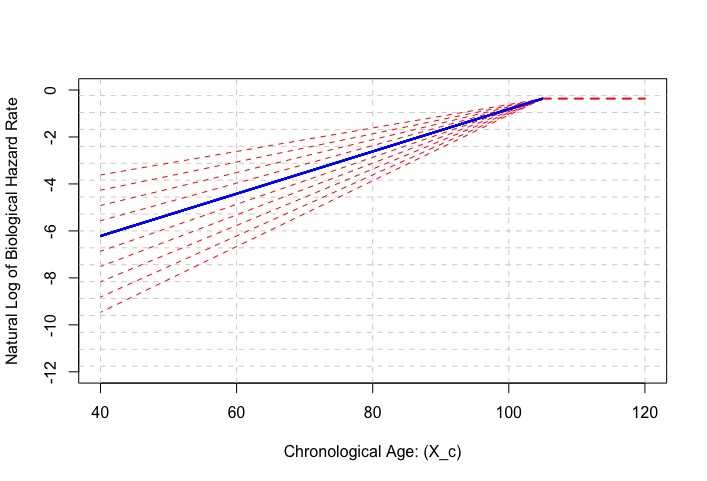} 
\includegraphics[width=0.45\textwidth]{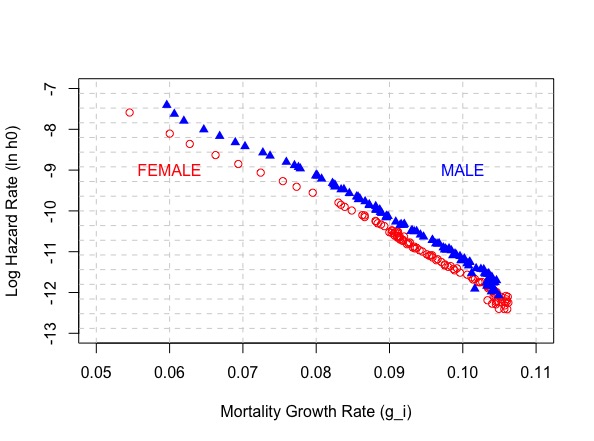} 

\end{center}
\end{figure}
Note: The {\em Compensation Law of Mortality} in its strong form implies that (log) mortality rates converge at some mortality plateau (and age) which then leads to a linear and negative relationship between intercept: $\ln h$, and slope: $g$, in the Gompertz regression. The right-side panel illustrates that relationship using the Chetty et al. (2016) data. 

\clearpage
\begin{figure}
\vspace{-0.2in}
\caption{The Coefficient of Variation of Longevity over the Lifecycle.}
\begin{center}
\includegraphics[width=0.95\textwidth]{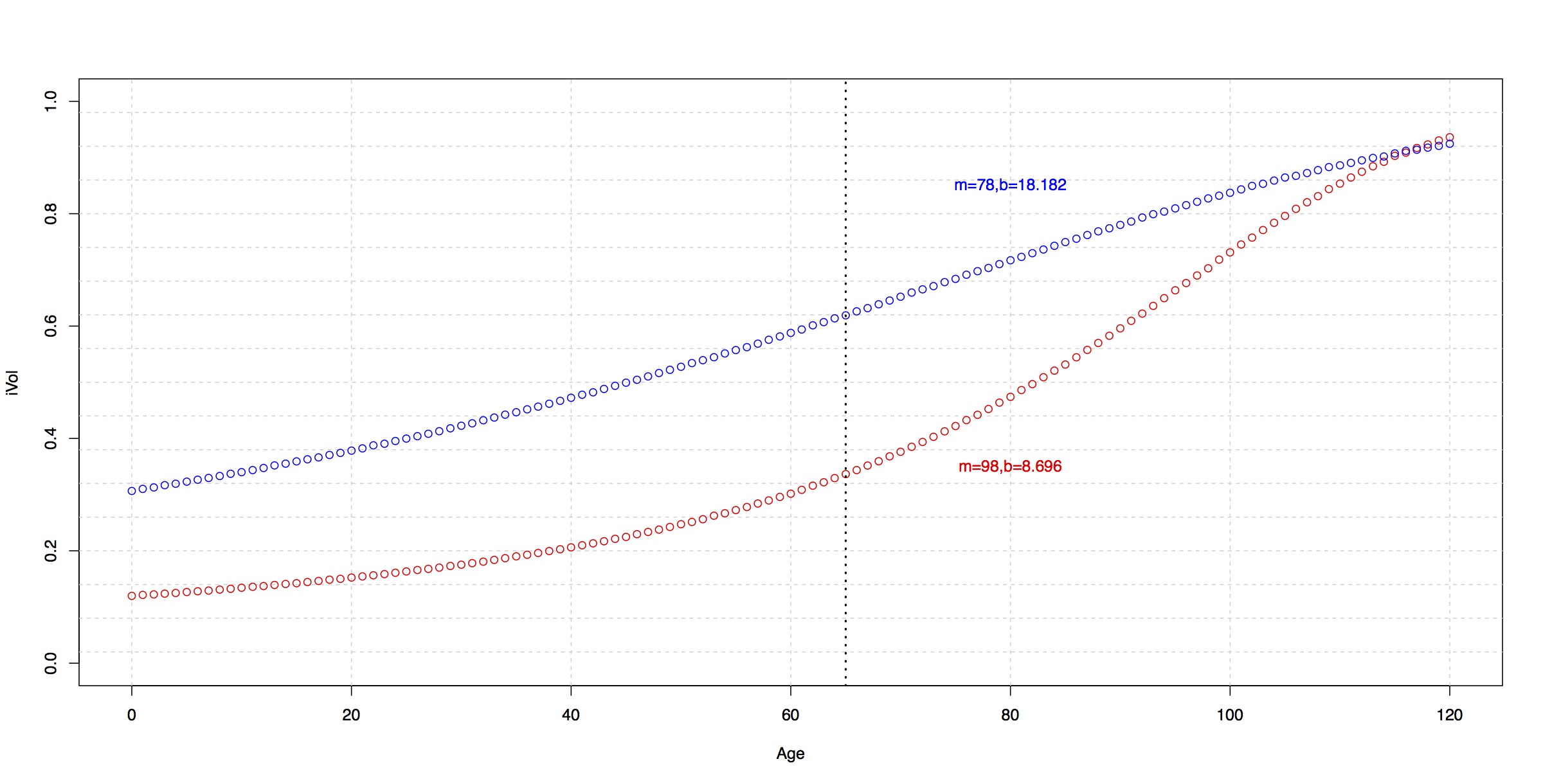} 
\end{center}
\end{figure}
Note: The y-axis (CoVoL) is defined as the ratio of the standard deviation of expected lifetime $SD[T_{x}]$ to mean lifetime $E[T_{x}]$ at age x. The vertical line is at age $x=65$, with exact values noted in table (\#3). The Gompertz parameters are: $m=98,b=8.696$ (rich) and $m=78,b=18.182$ (poor.) Notice how both curves and the CoVoL converge to a value of $\varphi=1$ at advanced ages independently of whether there is a mortality plateau.

\clearpage
\begin{figure}
\vspace{-0.2in}
\caption{Annuity Equivalent Wealth (AEW) under a range of values for $m$.}
\begin{center}
\includegraphics[width=0.95\textwidth]{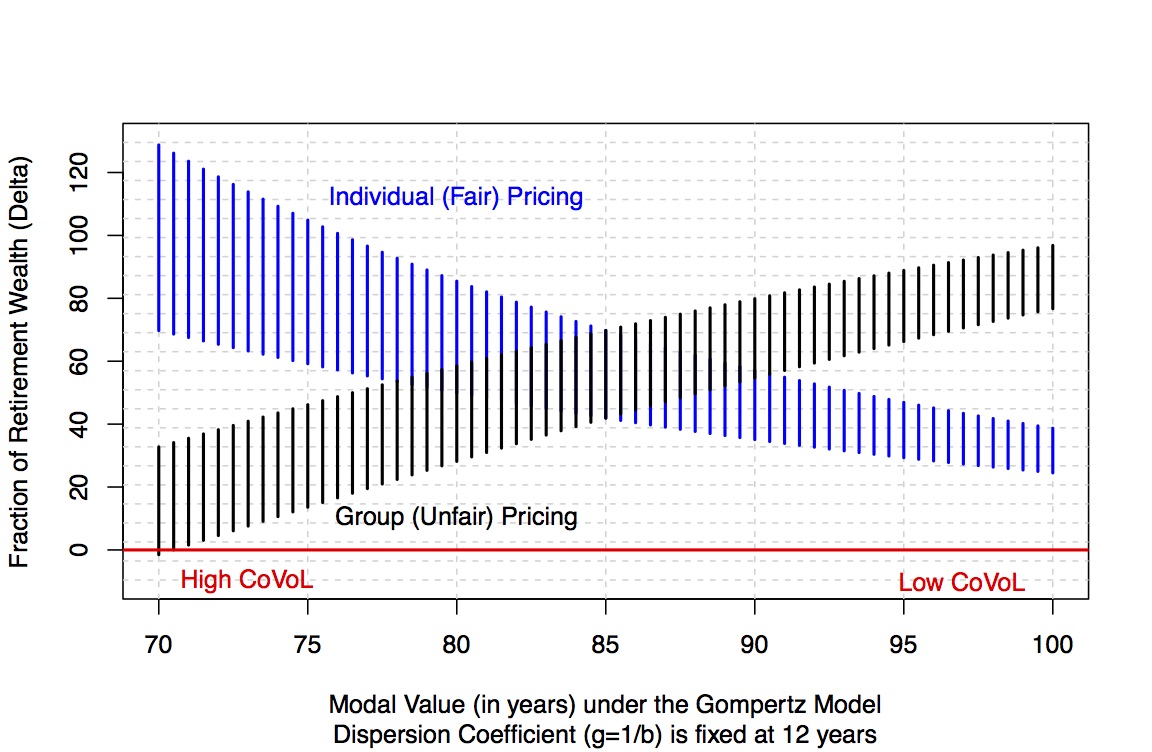} 
\end{center}
\end{figure}
Note: Figure is based on a valuation rate $r=3\%$ and coefficient of relative risk aversion from $\gamma=5$ (top) to $\gamma=1$ (bottom). When annuities are priced based on individual mortality $(m_i,b=12)$, the value of AEW declines in $m_i$. But when annuities are priced based on group mortality $(\hat{m},b=12)$, the value of AEW increases in $m_i$, due to the implicit loading.

\clearpage
\begin{figure}
\vspace{-0.2in}
\caption{Estimated Gompertz Parameter Values versus Income Percentiles}
\begin{center}
\includegraphics[width=0.95\textwidth]{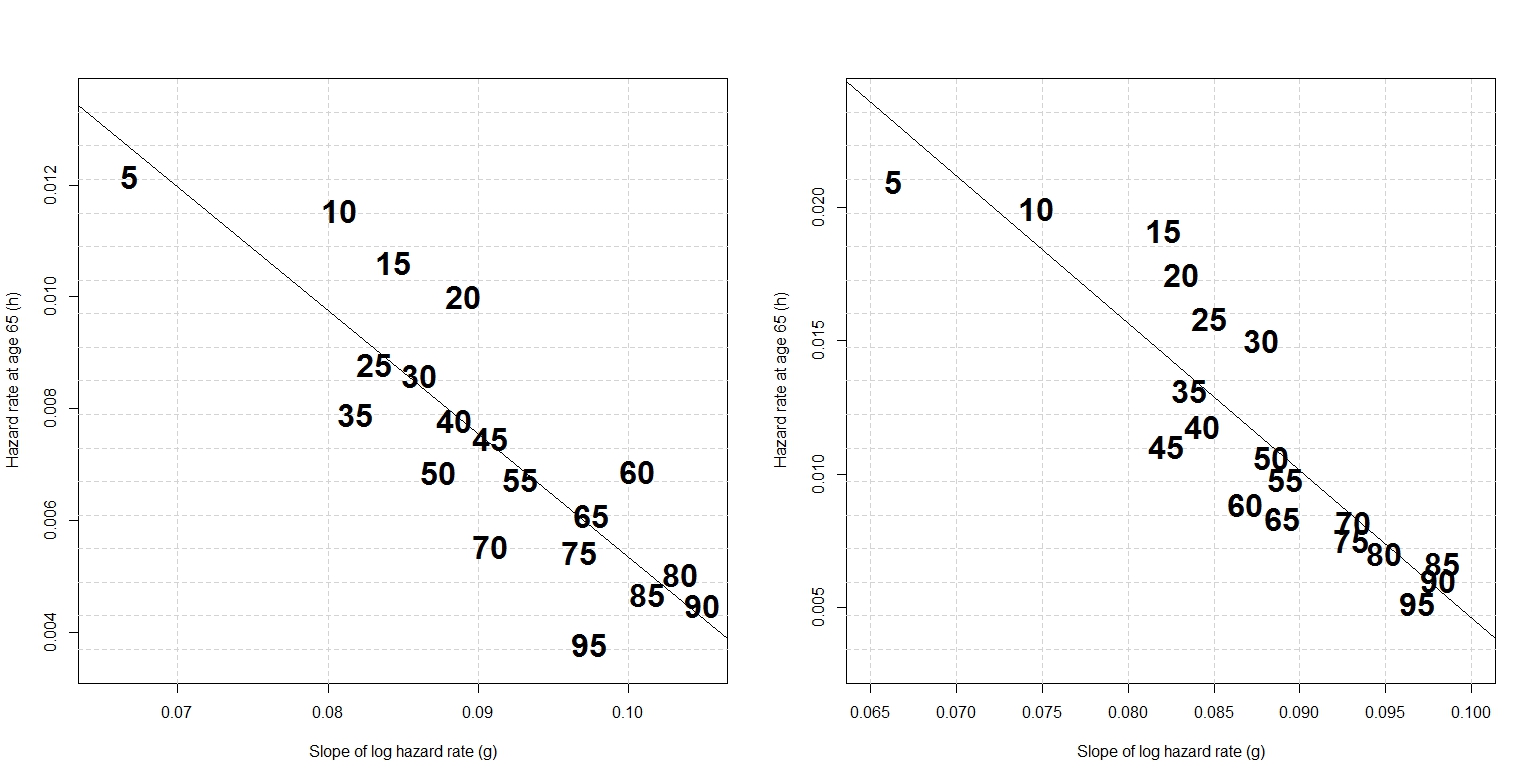} 
\end{center}
\end{figure}
Note: Using data from Chetty et al. (2016), average mortality rates during the 2001 to 2014 period are converted into Gompertz $(h,g)$ parameters for various income percentiles. Each percentile point noted in the figure (left panel for females and right for males) is placed at the co-ordinate for the relevant value.  Generally speaking higher (wealthier) income percentiles (numbers) are located at the bottom right and lower percentiles are at the top left. Wealthier retirees have lower mortality rates, but age faster. 

\clearpage
\begin{figure}
\vspace{-0.2in}
\caption{Canadian Evidence on the Compensation Law of Mortality}
\begin{center}
\includegraphics[width=0.95\textwidth]{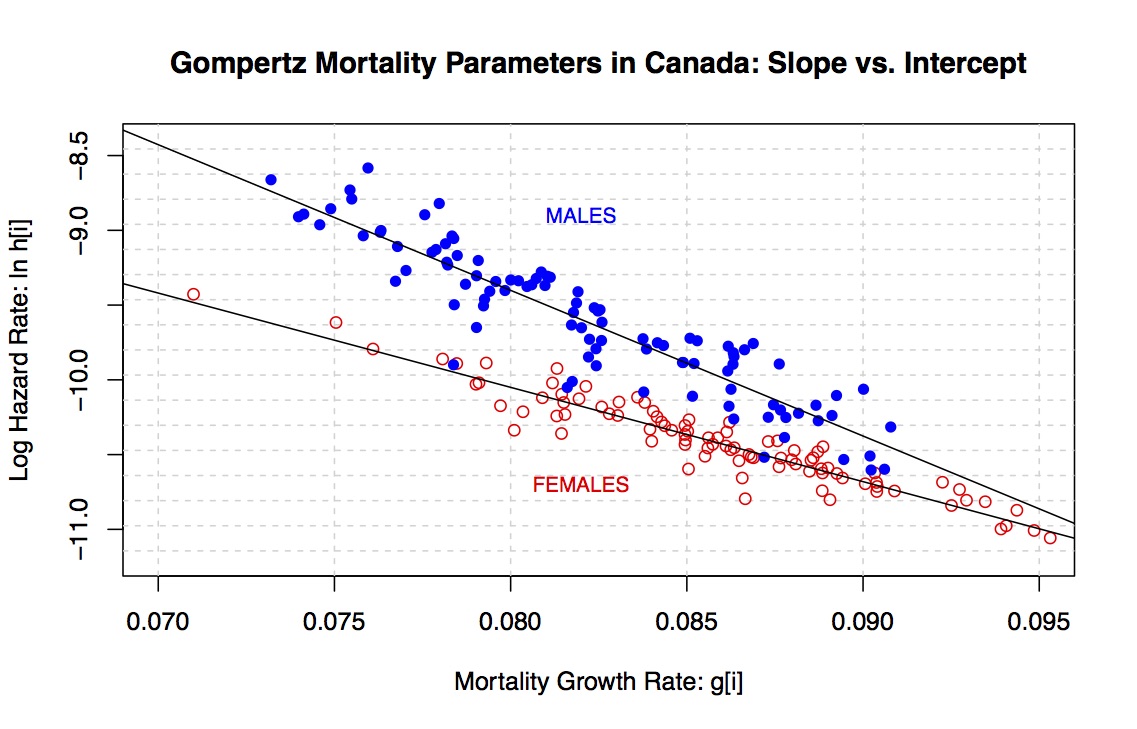} 
\end{center}
\end{figure}
Note: The Gompertz regression values ($\ln h,g$) were provided by Kevin Milligan, based on Canadian Pension Plan (CPP) data described and analyzed in Milligan and Schirle (2018). Similar to the Chetty et al. (2016) data, note the negative relationship between the age-zero (log) hazard rate and the mortality growth rate.

\clearpage
\begin{figure}
\vspace{-0.2in}
\caption{The Annuity Factor in $(h,g)$-space}
\begin{center}
\includegraphics[width=0.95\textwidth]{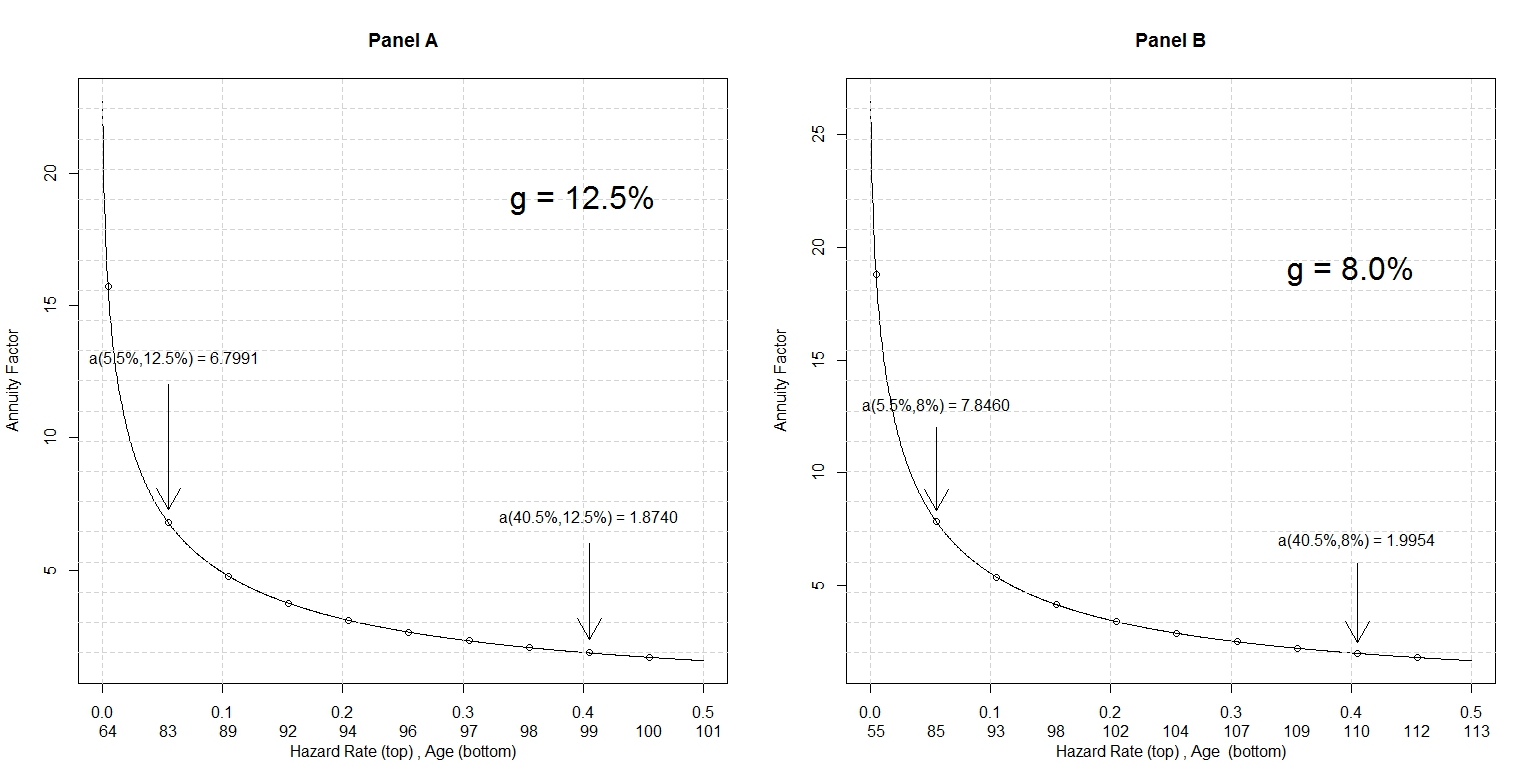} 
\end{center}
\end{figure}
Note: The cost of $\$1$ lifetime income is cheaper (and the annuity valuation factor is lower) when the current hazard rate $(h_x=h)$ is higher and/or the hazard growth rate $(g)$ is higher. Both are equivalent to increasing the discount rate $(r)$ in the present value factor. Technically; $\partial {\tt a}(r,h,g) / \partial r \leq 0$, $\partial {\tt a}(r,h,g) / \partial h \leq 0$, and: $\partial {\tt a}(r,h,g) / \partial g \leq 0$.

\clearpage
\begin{figure}
\vspace{-0.2in}
\caption{Comparative Statics: The value of $\delta_x$ as a function of $g$ on the CLaM line}
\begin{center}
\includegraphics[width=0.95\textwidth]{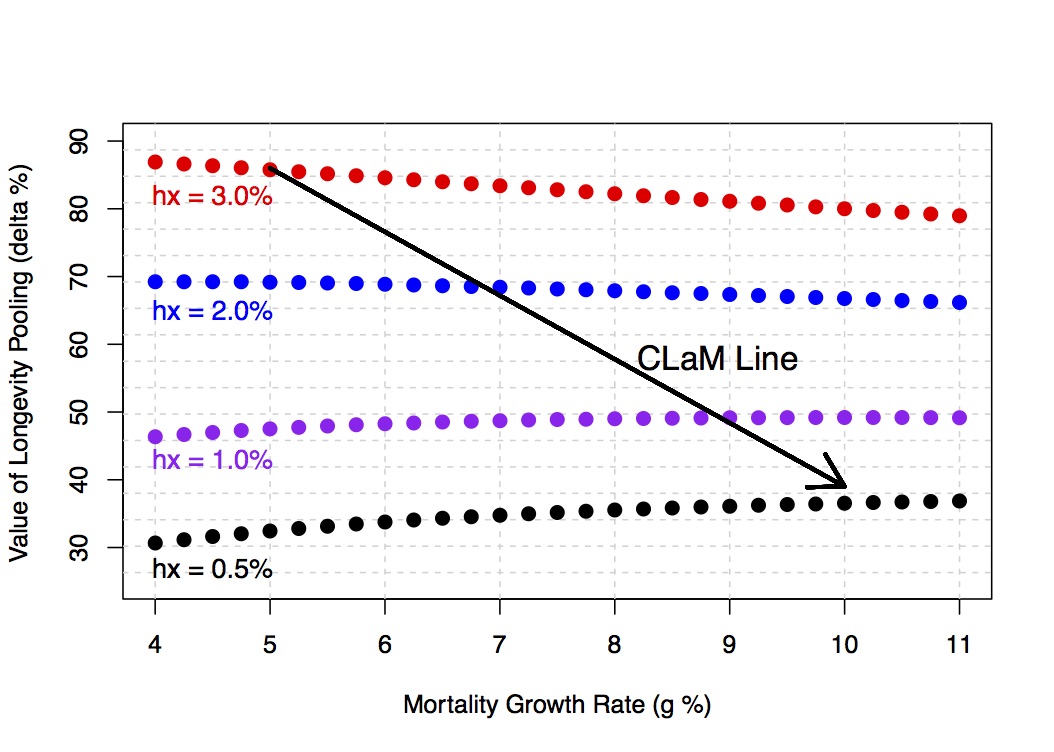} 
\end{center}
\end{figure}
Note: This assumes $r=3\%$ and the coefficient of relative risk aversion: $\gamma=3$.

\end{document}